\documentclass[preprint,review,12pt]{elsarticle}

\usepackage{amssymb}

\newcommand{\fr}[2]{{\textstyle \frac{#1}{#2}}}
\newcommand{\kp}{\varkappa}
\newcommand{\kpa}{\kp_a}

\newcommand{\ex}[1]{{\rm e}^{#1}}
\newcommand{\sh}{{\rm sh}}
\newcommand{\ch}{{\rm ch}}
\newcommand{\qref}[1]{(\ref{#1})}
\newcommand{\ra}{$\rightarrow\;$}
\newcommand{\vol}[1]{{#1}}

\biboptions{square,sort&compress}

\journal{Physica B}

\begin{document}

\begin{frontmatter}



\title{Phase diagrams of spin-3/2 Blume-Capel model on rectangular lattice
under longitudinal magnetic field}


\author{O. Baran\corref{cor1}}
\ead{ost@icmp.lviv.ua}
\cortext[cor1]{Corresponding author}

\author{R. Levitskii}

\address{Institute for Condensed Matter Physics of the National Academy of
Sciences of Ukraine, 1 Svientsitskii Str., 79011 L'viv, Ukraine}

\begin{abstract}
The spin-3/2 Blume-Capel model on a rectangular lattice
with the ferromagnetic bilinear short-range interaction
$K^{\rm F}=K(1+x)$ in one direction and the anti-ferromagnetic one
$K^{\rm AF}=K(-1+x)$ in the perpendicular direction under a longitudinal
magnetic field and in the presence of a single-ion anisotropy is
investigated within the mean field approximation. The quantity $x$
characterizes coupling between the interactions in the
perpendicular directions and its absolute value does not exceed
unity.
Basing on the study of temperature dependencies of sublattice
magnetizations the phase diagrams in the ($x$, temperature)
plane are constructed for different values of the
longitudinal magnetic field and the single-ion anisotropy.
The global phase diagram of the model demonstrates an unusual abundance.

\end{abstract}


\end{frontmatter}


\section{Introduction}

The Ising models with spin $S>1/2$ and their numerous variations have
been extensively studied during recent years due to a
relative simplicity with which calculations for these models
within different techniques can be carried out for
testing of these methods as well as  the theoretical interest
arising from the richness of the phase diagrams
due to competition of models parameters
\cite{Takahashi,Kasono,Netz,Bakchich,
Bekhechi,Baran1,Baran2,Baran3,Ekiz1,Ekiz2,Keskin1,Keskin2}.
On the other hand, most modifications
of the Ising model are applied for description of
real objects. Namely, the spin-3/2 Blume-Emery-Griffiths model with
nearest-neighbor interactions, both bilinear and biquadratic, and
a single-ion anisotropy (or a crystal-field interaction) was
introduced to explain the phase transition in DyVO$_4$ qualitatively
\cite{Sivardiere2}  and proved to be useful for description
tricritical properties of ternary fluid mixtures \cite{Krinsky}.
The spin-3/2 Blume-Capel model, which is the partial case of
the spin-3/2 Blume-Emery-Griffiths model without
biquadratic interactions, can be applied to study KEr(MoO$_4$)$_2$
\cite{Horvath,Orendacova}.

Theoretical techniques used to investigate
the spin-3/2 Ising models include the mean field approximation (MFA)
\cite{Bakchich,Keskin1,Keskin2,Sivardiere2,Krinsky,bor1,Baretto},
the effective field theory \cite{Bakkali,Kaneyoshi},
the cluster approximation as well as the Bethe approximation
(the exact results for Bethe lattices)
\cite{Ekiz1,Ekiz2,Tucker},
the renormalization-group method \cite{Bakchich2},
the transfer-matrix finite-size-scaling calculations \cite{Bekhechi}.
These models have also been studied by means of the
Monte-Carlo simulations
\cite{Bekhechi,Baretto,Xavier}.

Most of the papers, where the spin-3/2 Ising models are investigated,
consider the lattices with either ferromagnetic bilinear interactions or
antiferromagnetic ones.
In this work we study the spin-3/2 Blume-Capel model
\begin{eqnarray}
\label{f1} &&
H = - \sum_{i=1}^L \sum_{j=1}^L \Bigg[ \Gamma
S_{i,j} + D S_{i,j}^2 + K^{\rm F}  S_{i,j} S_{i+1,j} + K^{\rm AF}  S_{i,j}
S_{i,j+1} \Bigg]
\end{eqnarray}
on the rectangular lattice with the
ferromagnetic bilinear short-range interaction $K^{\rm F}=K(1+x)$ in
one direction and the antiferromagnetic one $K^{\rm AF}=K(-1+x)$ in
the perpendicular direction,
where $\Gamma$ is a longitudinal magnetic field, $D$ is a single-ion anisotropy,
$K>0$, $x\in]-1,1[$. Each $S_{i,j}=S_{i,j}^z$ takes the value
$\pm 1/2$ or $\pm 3/2$.
We will especially investigate the effect of swapping of interactions
strength $x$ within the MFA.

\section{Mean field approximation}
The expression for a free energy of the Blume-Capel model
(\ref{f1}) within the MFA is constructed on
the basis of the single-site Hamiltonians
\begin{eqnarray}
\label{f2} && \hspace{-10mm}
H_{i_a} = - \kpa S_{i_a} - D S^2_{i_a} ,
\\ \label{f2a} && \hspace{-10mm}
\kpa=\Gamma + 2 K^{\rm F} m_a + 2 K^{\rm AF} m_b,
\qquad
 m_a=\langle S_{i_a} \rangle , \qquad
(a,b = A,B)
\end{eqnarray}
(where $A$ and $B$ refer to two different sublattices)
in a usual way
\cite{bor1,Smart,Chen,Blume}
as follows:
\begin{eqnarray} \label{f3} && \hspace{-10mm}
F=-\fr{N}{2} k_{\rm B} T  \Bigg[  {\rm ln} Z_{1_A} + {\rm ln} Z_{1_B} \Bigg]
+\fr{N}{2}K^{\rm F}(m_A^2+m_B^2) + N K^{\rm AF} m_A m_B ,
\\ \label{f4} && \hspace{-10mm}
Z_{1_a}= 2 \ex{\beta D /4} \Bigg[ \ch(\beta \kpa / 2) +
\ex{2 \beta D } \ch(3 \beta \kpa / 2) \Bigg] .
\end{eqnarray}
Here $\beta=(k_{\rm B}T)^{-1}$, and $N=L^2$ is the total number of spins.

Within the MFA the
magnetizations of sublattices $m_A$ and $m_B$ can be
determined \cite{Orendacova,bor1,Smart,Chen,Blume,Yoshizawa}
from the conditions of extremum of the free energy
with respect to them ($\partial F / \partial m_A=\partial F / \partial m_B=0$).
These conditions yield the following system of equations for $m_a$:
\begin{eqnarray}
\label{f5} && \nonumber
\ex{\beta D /4} \Bigg[ \sh(\beta \kp_A / 2) +
3 \ex{2 \beta D } \sh(3 \beta \kp_A / 2) \Bigg] / Z_{1_A}  - m_A = 0 ,
\\[12pt] &&
\ex{\beta D /4} \Bigg[ \sh(\beta \kp_B / 2) +
3 \ex{2 \beta D } \sh(3 \beta \kp_B / 2) \Bigg] / Z_{1_B}  - m_B = 0 .
\end{eqnarray}

\section{Numerical analysis results}

Let us consider results of our numerical investigation of
model \qref{f1} within the MFA.

To gain a better insight into the following results, let us explain principles
of construction of phase diagrams and classification of phases.
At given values of model parameters
$K$, $x$ (i.e. fixed values of $K^{\rm F}=K(1+x)$ and $K^{\rm AF}=K(-1+x)$),
$\Gamma$, and $D$ we study temperature dependencies of solutions of the
equation set \qref{f5}.
The number of these solutions (i.e. pairs of temperature dependent branches
of solutions)
varies at different values of model parameters.
Some pairs of branches-solutions exist on the whole temperature interval
$T \in [0,\infty[$, others are defined on the limited intervals starting from the
zero temperature $T \in [0,T_1]$, or the non-zero one $T \in [T_2,T_3]$.
Choosing pairs of solutions which correspond to the absolute minimum of the free
energy, we obtain temperature dependencies for magnetizations of sublattices.
Calculating in this manner sublattice magnetizations for ranges of values of model
parameters we construct phase diagrams for the considered model.

At temperature of the first order phase transition magnetizations of
sub\-lattices ``jump'' from one pair of branches-solutions of the
equation set \qref{f5} to other one.
Thus, different thermodynamically stable pairs of branches-solutions
correspond to different phases (on the respective temperature intervals).
Names of the ordered phases include values of solutions at the zero temperature
which corresponds to the branches-solutions describing sublattice magnetizations
in the chosen phase.

Thus we shall distinguish the following phases (see \cite{Bakchich}):

\noindent
$\bullet$ disordered phase ({\bf d});

\noindent
$\bullet$ antiferrimagnetic phase with $m_A=-3/2$, $m_B=3/2$ at $T=0$
({\bf ai}$_{-3/2,+3/2}$);

\noindent
$\bullet$ antiferrimagnetic phase with $m_A=-1/2$, $m_B=3/2$ at $T=0$
({\bf ai}$_{-1/2,+3/2}$);

\noindent
$\bullet$ antiferrimagnetic phase with $m_A=-1/2$, $m_B=1/2$ at $T=0$
({\bf ai}$_{-1/2,+1/2}$);

\noindent
$\bullet$ ferrimagnetic phase with $m_A=1/2$, $m_B=3/2$ at $T=0$
({\bf i}$_{+1/2,+3/2}$).

However, in the considered system there are also ordered phases where sublattice
magnetizations are defined by the branches-solutions starting from non-zero temperatures.
Such phases are simply designed as ferrimagnetic ({\bf i}) or antiferromagnetic
({\bf ai}) without indices describing ground state magnetizations.

This simple and clear classification of phases by indices, corresponding to the
magnetizations in the ground state, is convenient for investigation of temperature
dependencies of sublattice magnetizations because at $T=0$ magnetizations reach
their asymptotical values (i.e.
$m_A=-3/2$, $m_B=3/2$, or $m_A=-1/2$, $m_B=3/2$, or
$m_A=-1/2$, $m_B=1/2$, or $m_A=1/2$, $m_B=3/2$).
However, there are also phase transitions between ordered phases at variation
of the $x$ parameter.
While studying $x$-dependencies of magnetizations
at $T>0$, such ordered phases can not be designed by
the same indices as we do it considering temperature dependencies.

Let us also note that within MFA the phase transitions
between different ordered phases
can be only of the first order and
the transitions between ordered and disordered phases can be both of
the second and of the first order.

\begin{figure}[t]
\centerline{\includegraphics[width=0.62\textwidth]{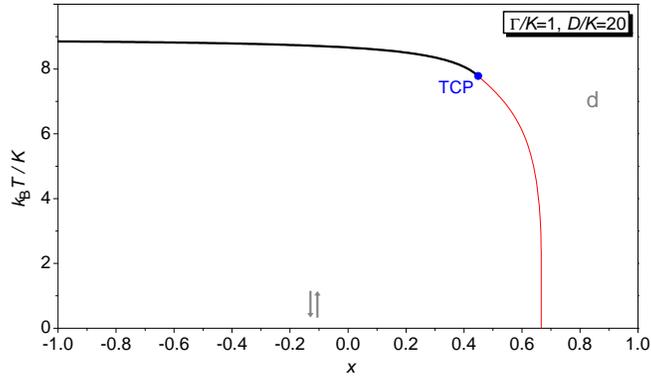}}
\caption{
The $T$ vs $x$ phase diagram at $D/K=20$ and $\Gamma /K=1$.
Thick  and thin lines correspond to the phase transition between
ordered and disordered phases of the second and the first order, respectively.
The special point is a tricritical point.
} \label{fig01}
\end{figure}

\begin{figure}[b]
\centerline{\includegraphics[width=0.62\textwidth]{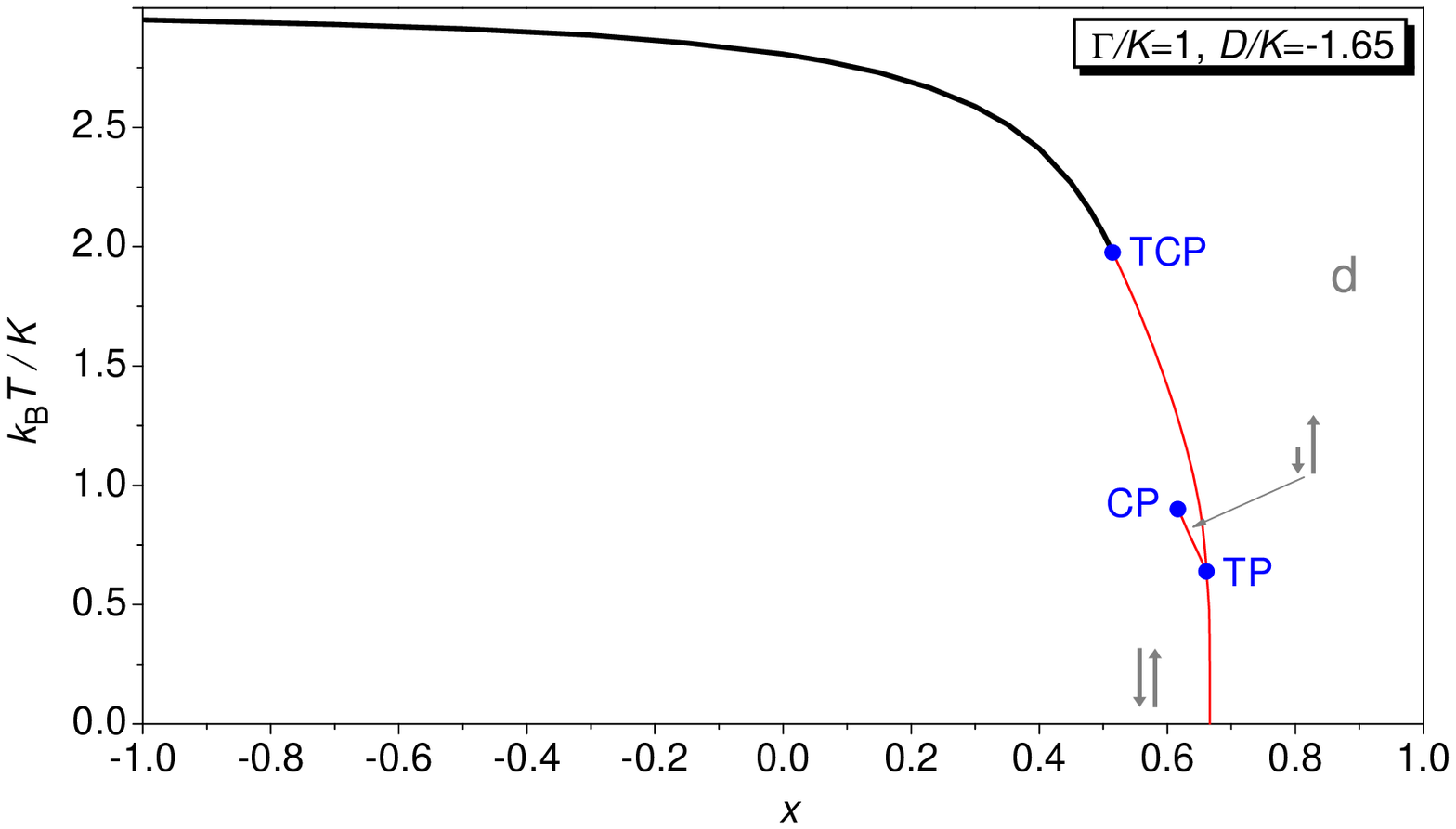}}
\caption{
The $T$ vs $x$ phase diagram at $D/K=-1.65$ and $\Gamma /K=1$.
A thick line indicates the
phase transition between ordered and disordered phases of the second order.
Thin lines indicate the first order phase
transitions between ordered and disordered phases
as well as between different ordered phases.
The special points are a tricritical point, a critical point and a triple point.
} \label{fig02}
\end{figure}

\begin{figure}[h]
\centerline{\includegraphics[width=0.62\textwidth]{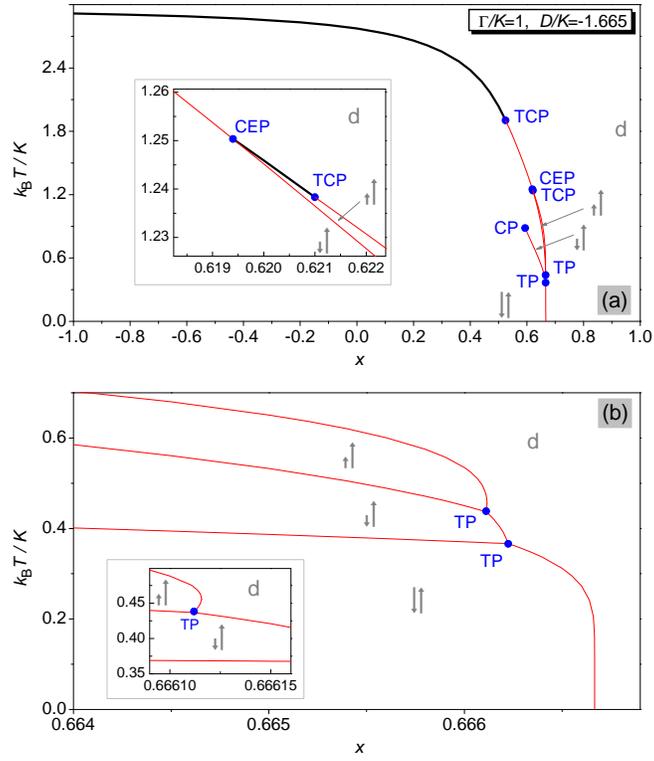}}
\caption{
The $T$ vs $x$ phase diagram at $D/K=-1.665$ and $\Gamma /K=1$.
A thick line indicates the
phase transition between ordered and disordered phases of the second order.
Thin lines indicate the first order phase
transitions between ordered and disordered phases
as well as between different ordered phases.
The special points are tricritical points, a critical end point, a critical point
and triple points.
Enlarge parts of the phase diagram are presented in subfigure (b) and
inserts.
} \label{fig03}
\end{figure}

\begin{figure}[h]
\centerline{\includegraphics[width=0.62\textwidth]{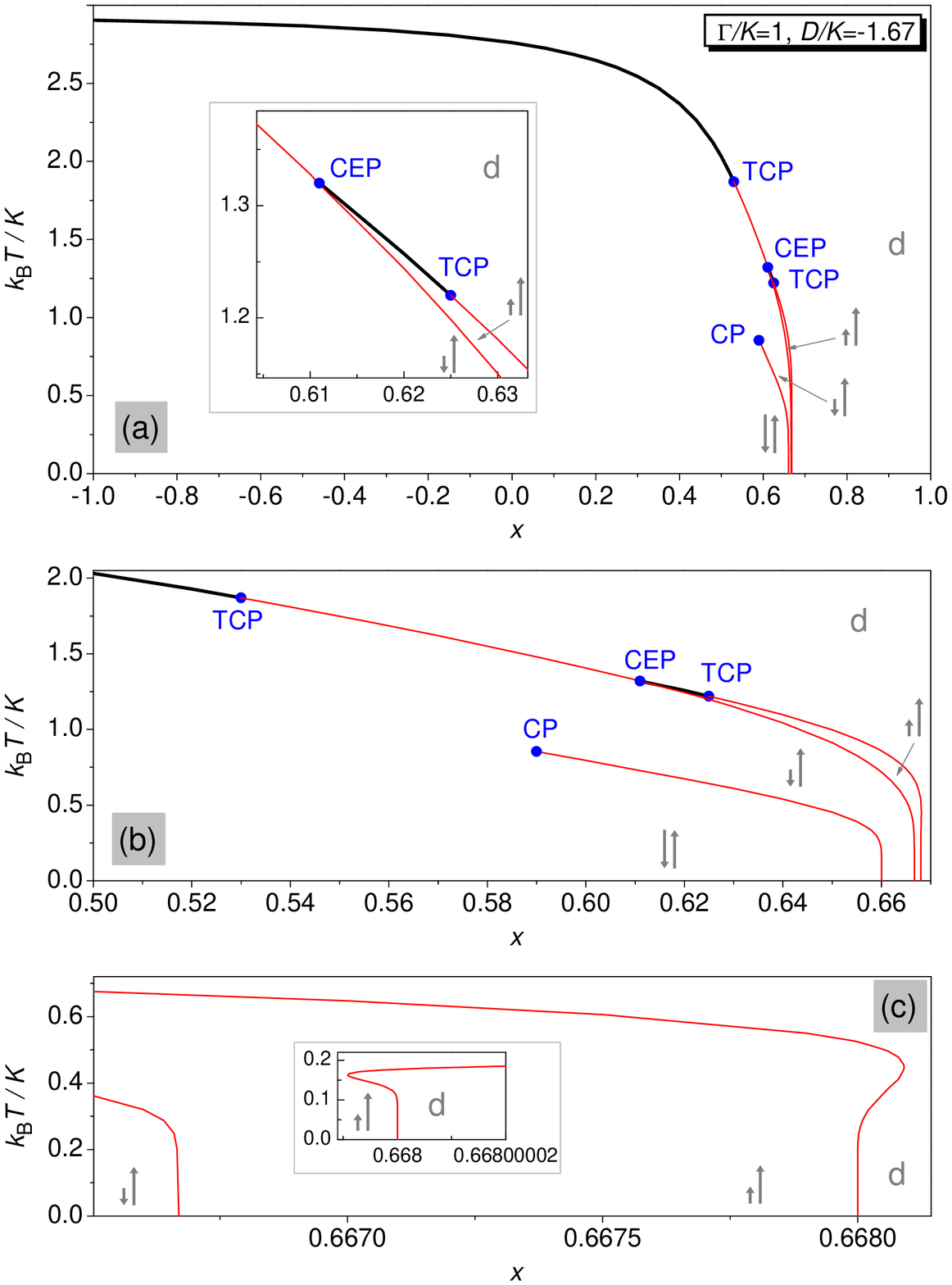}}
\caption{
The same as in Fig. \ref{fig03}, but $D/K=-1.67$ and $\Gamma /K=1$.
} \label{fig04}
\end{figure}

\begin{figure}[h]
\centerline{\includegraphics[width=0.62\textwidth]{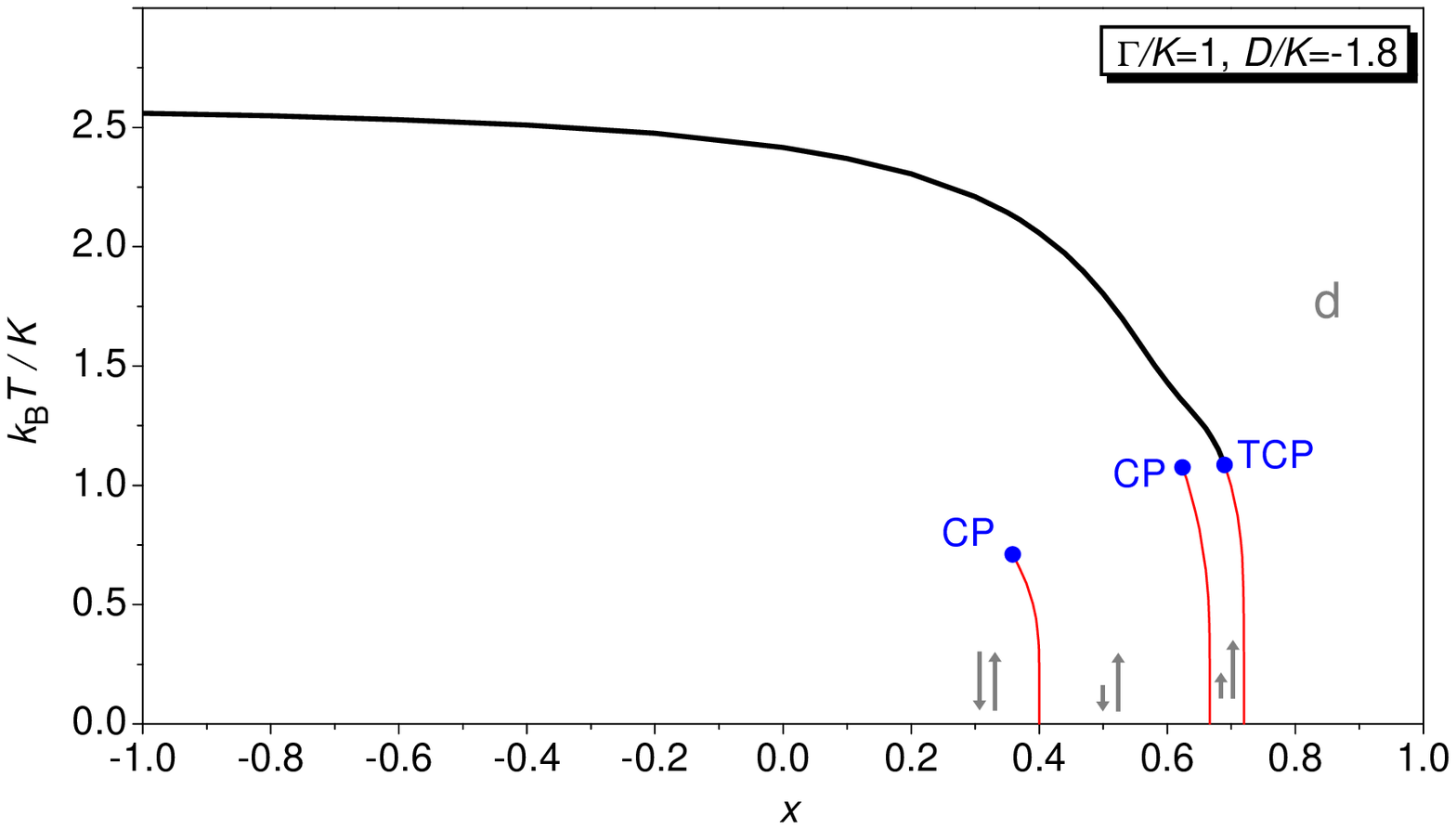}}
\caption{
The same as in Fig. \ref{fig03}, but $D/K=-1.8$ and $\Gamma /K=1$.
} \label{fig05}
\end{figure}

\begin{figure}[h]
\centerline{\includegraphics[width=0.62\textwidth]{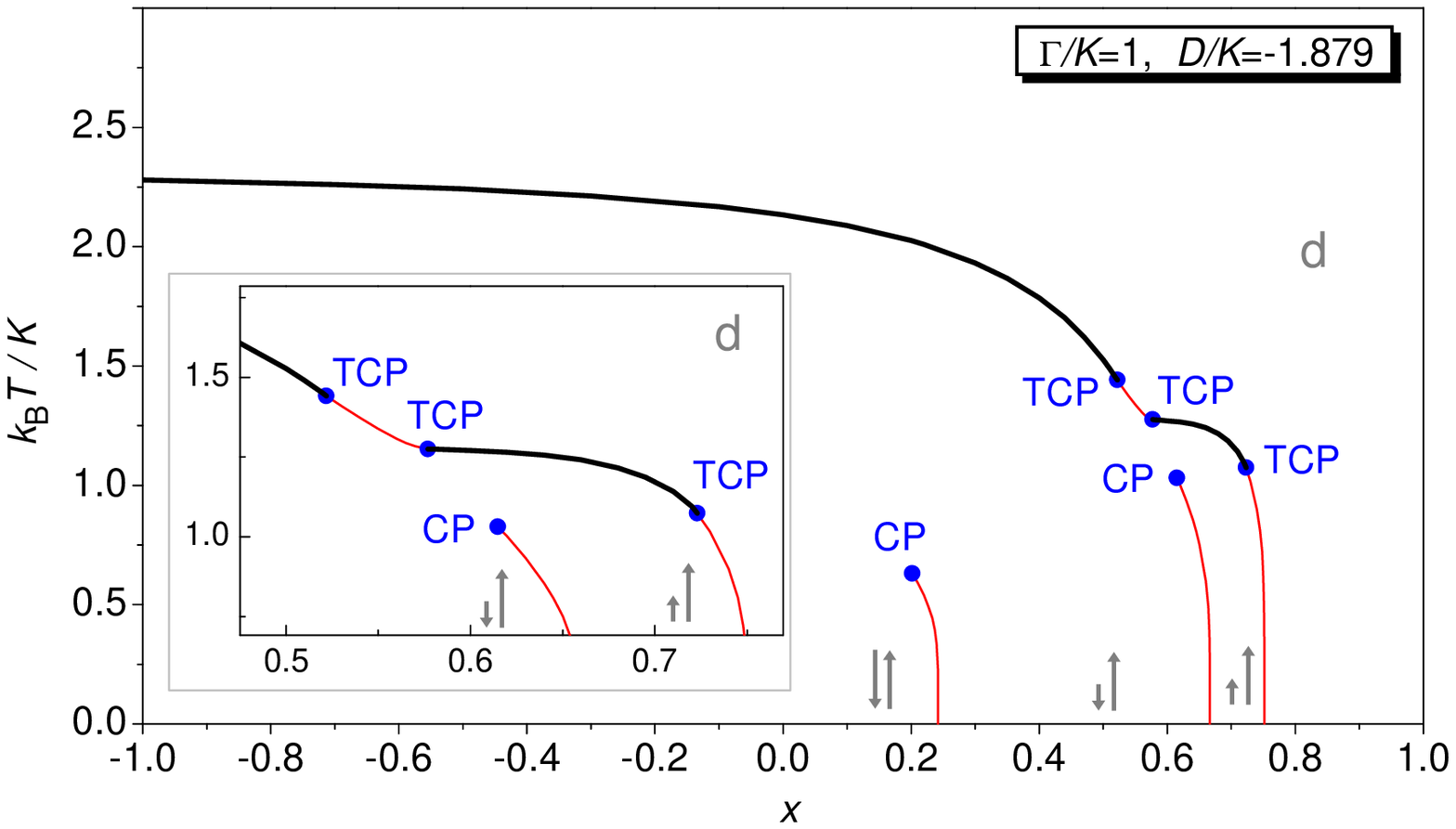}}
\caption{
The same as in Fig. \ref{fig03}, but $D/K=-1.879$ and $\Gamma /K=1$.
} \label{fig06a}
\end{figure}

\begin{figure}[h]
\centerline{\includegraphics[width=0.62\textwidth]{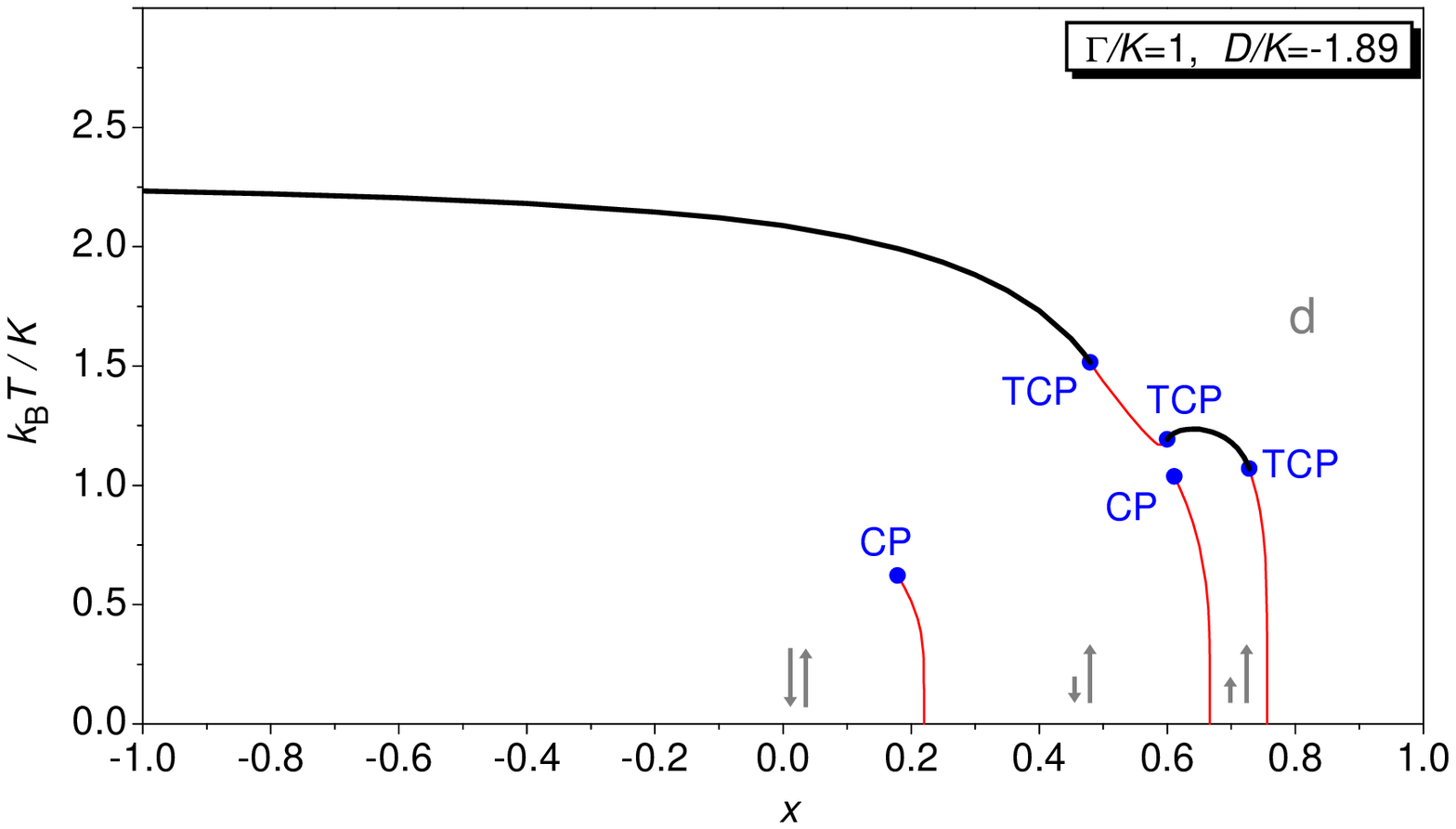}}
\caption{
The same as in Fig. \ref{fig03}, but $D/K=-1.89$ and $\Gamma /K=1$.
} \label{fig06}
\end{figure}

\begin{figure}[h]
\centerline{\includegraphics[width=0.62\textwidth]{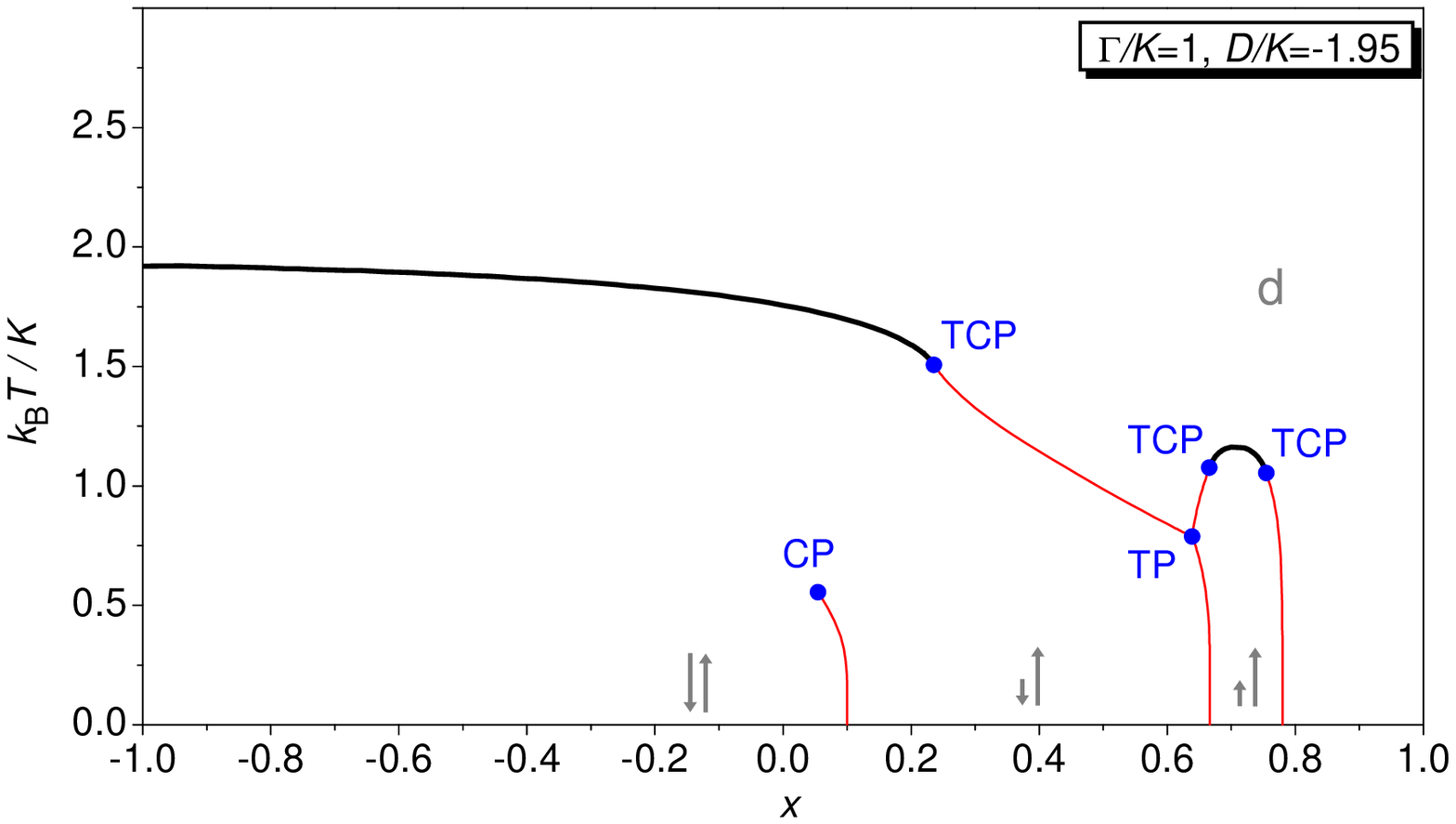}}
\caption{
The same as in Fig. \ref{fig03}, but $D/K=-1.95$ and $\Gamma /K=1$.
} \label{fig07}
\end{figure}

\begin{figure}[h]
\centerline{\includegraphics[width=0.62\textwidth]{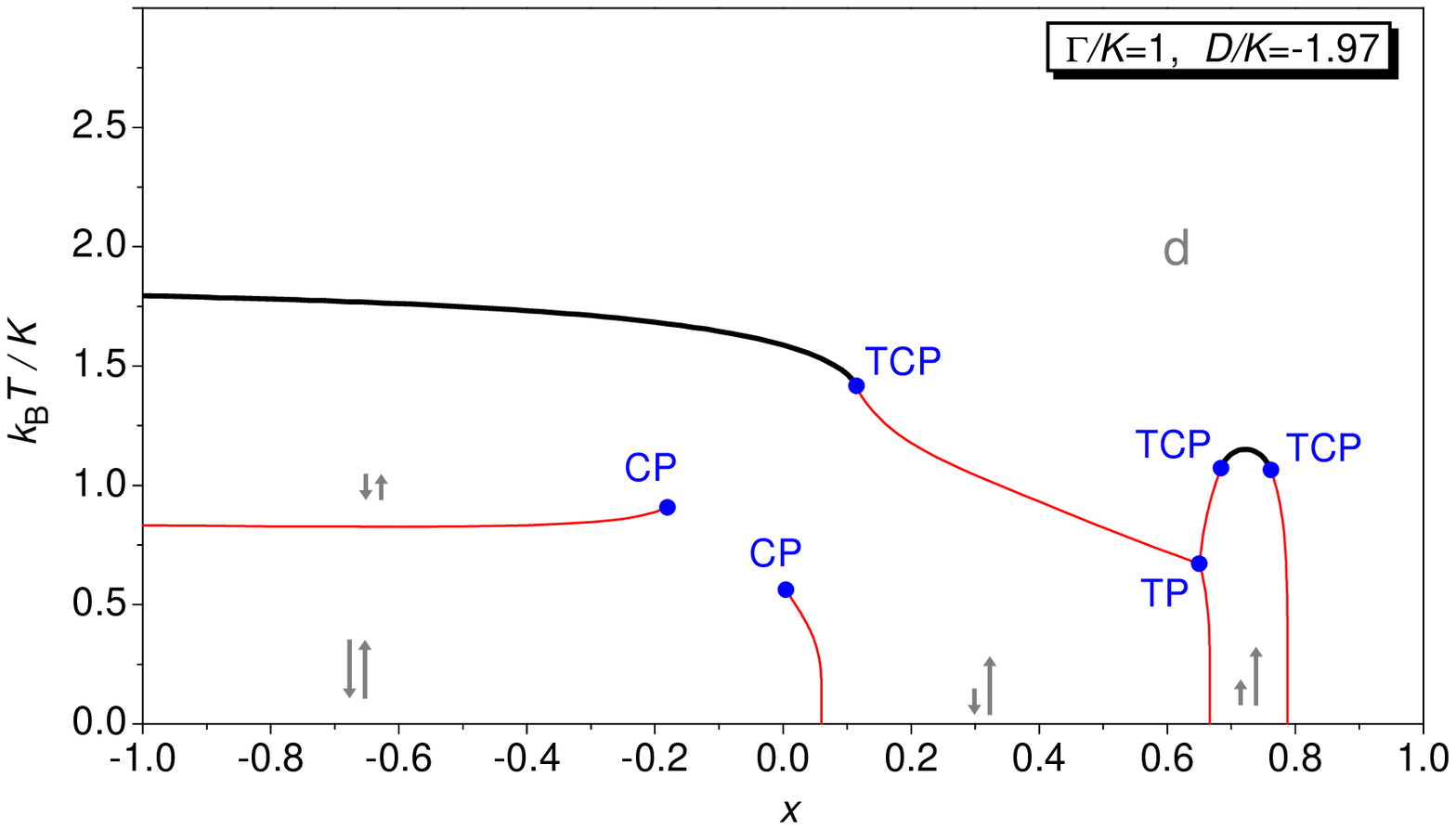}}
\caption{
The same as in Fig. \ref{fig03}, but $D/K=-1.97$ and $\Gamma /K=1$.
} \label{fig08}
\end{figure}

\begin{figure}[h]
\centerline{\includegraphics[width=0.62\textwidth]{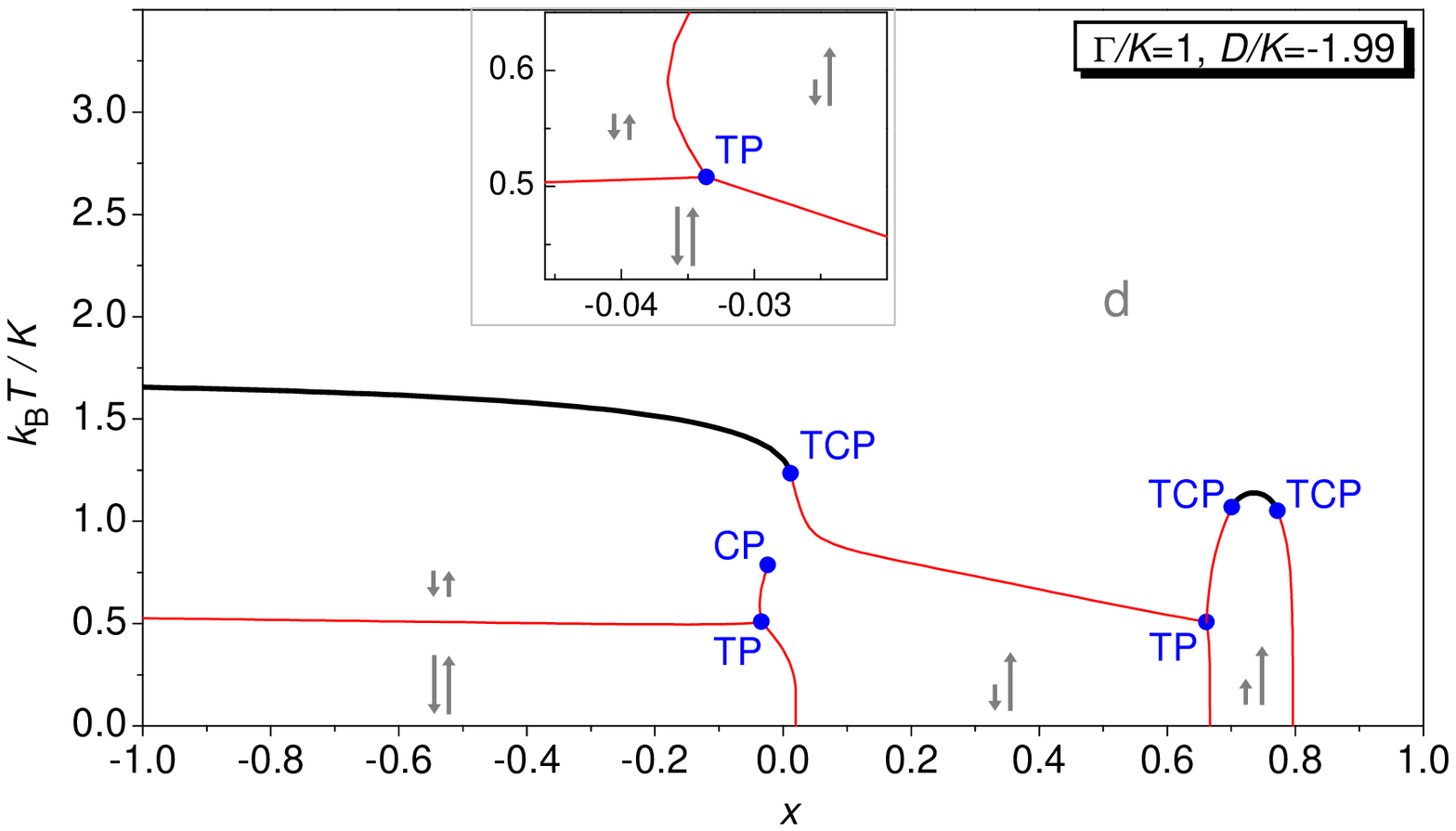}}
\caption{
The same as in Fig. \ref{fig03}, but $D/K=-1.99$ and $\Gamma /K=1$.
} \label{fig09}
\end{figure}

\begin{figure}[h]
\centerline{\includegraphics[width=0.62\textwidth]{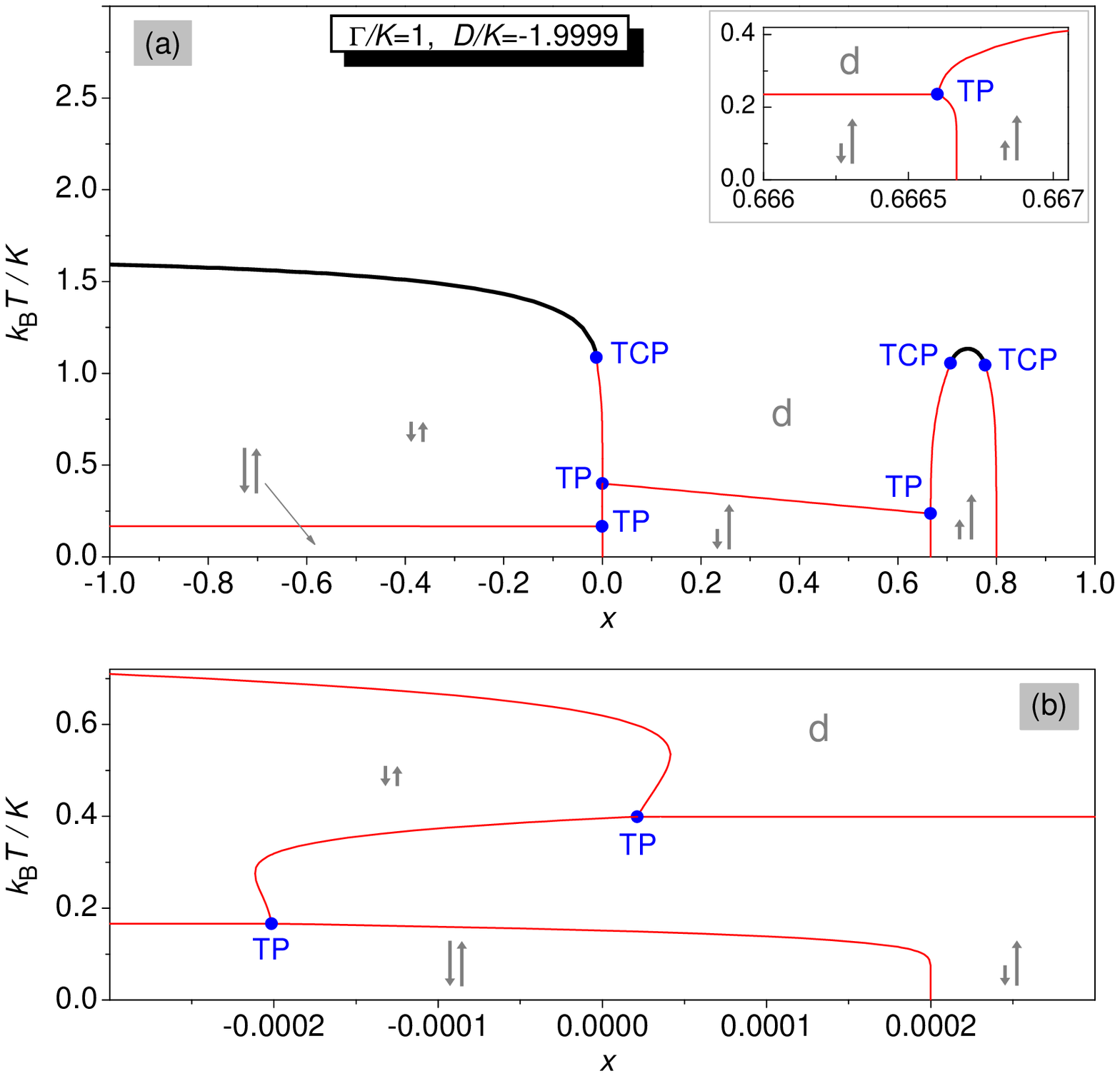}}
\caption{
The same as in Fig. \ref{fig03}, but $D/K=-1.9999$ and $\Gamma /K=1$.
} \label{fig10}
\end{figure}

\begin{figure}[h]
\centerline{\includegraphics[width=0.62\textwidth]{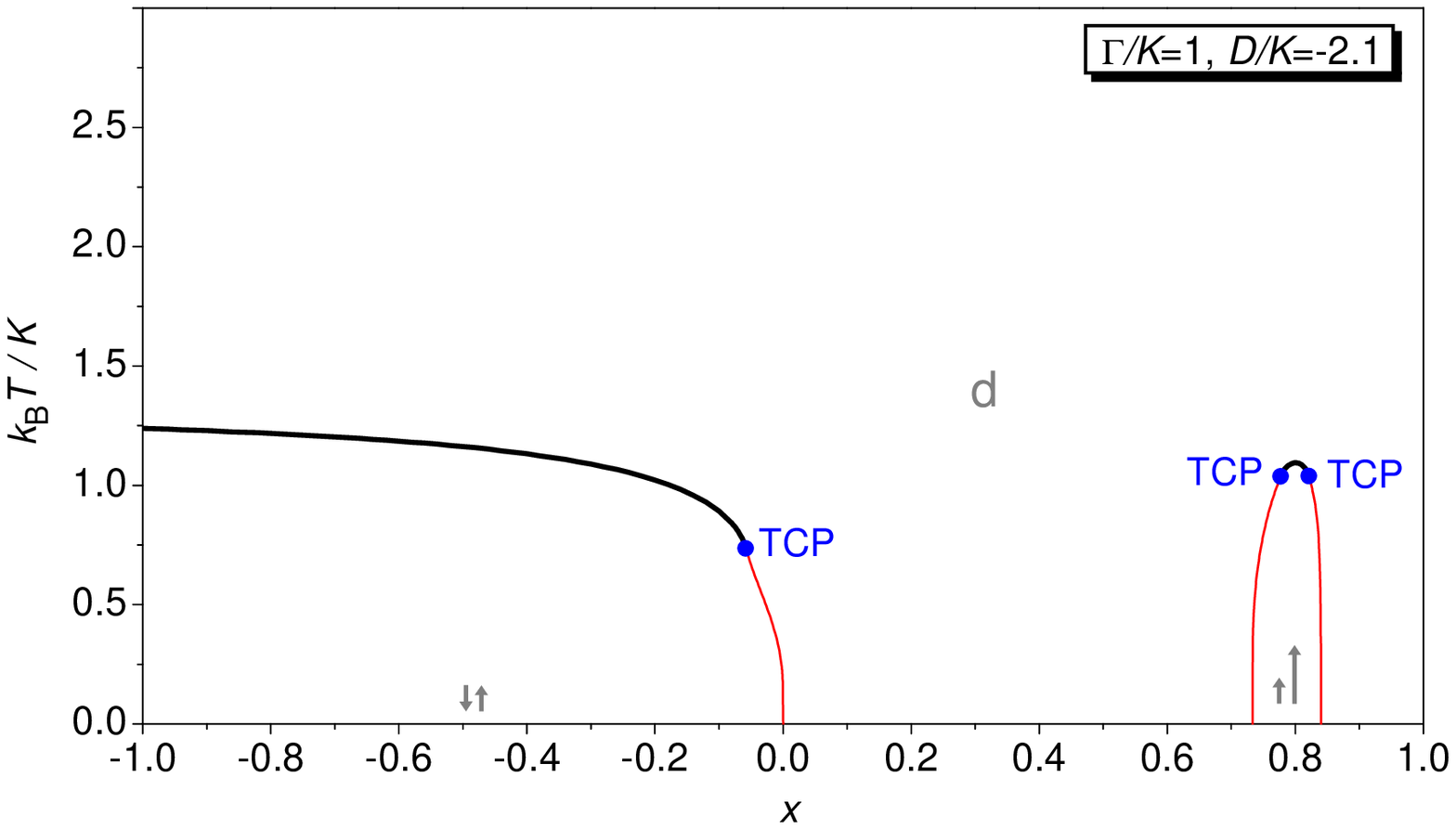}}
\caption{
The same as in Fig. \ref{fig03}, but $D/K=-2.1$ and $\Gamma /K=1$.
} \label{fig11}
\end{figure}

\begin{figure}[h]
\centerline{\includegraphics[width=0.62\textwidth]{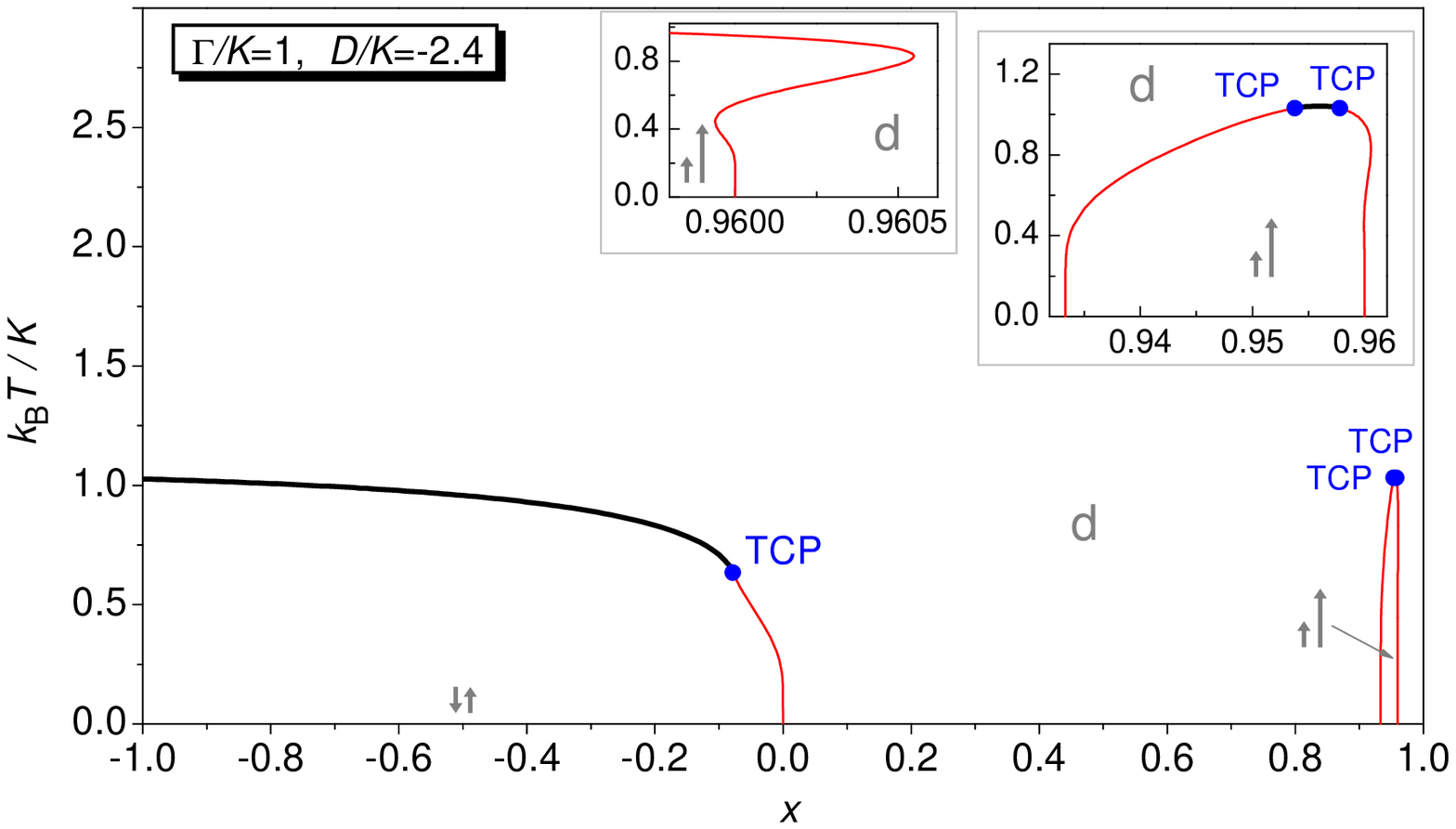}}
\caption{
The same as in Fig. \ref{fig03}, but $D/K=-2.4$ and $\Gamma /K=1$.
} \label{fig11b}
\end{figure}

\begin{figure}[h]
\centerline{\includegraphics[width=0.62\textwidth]{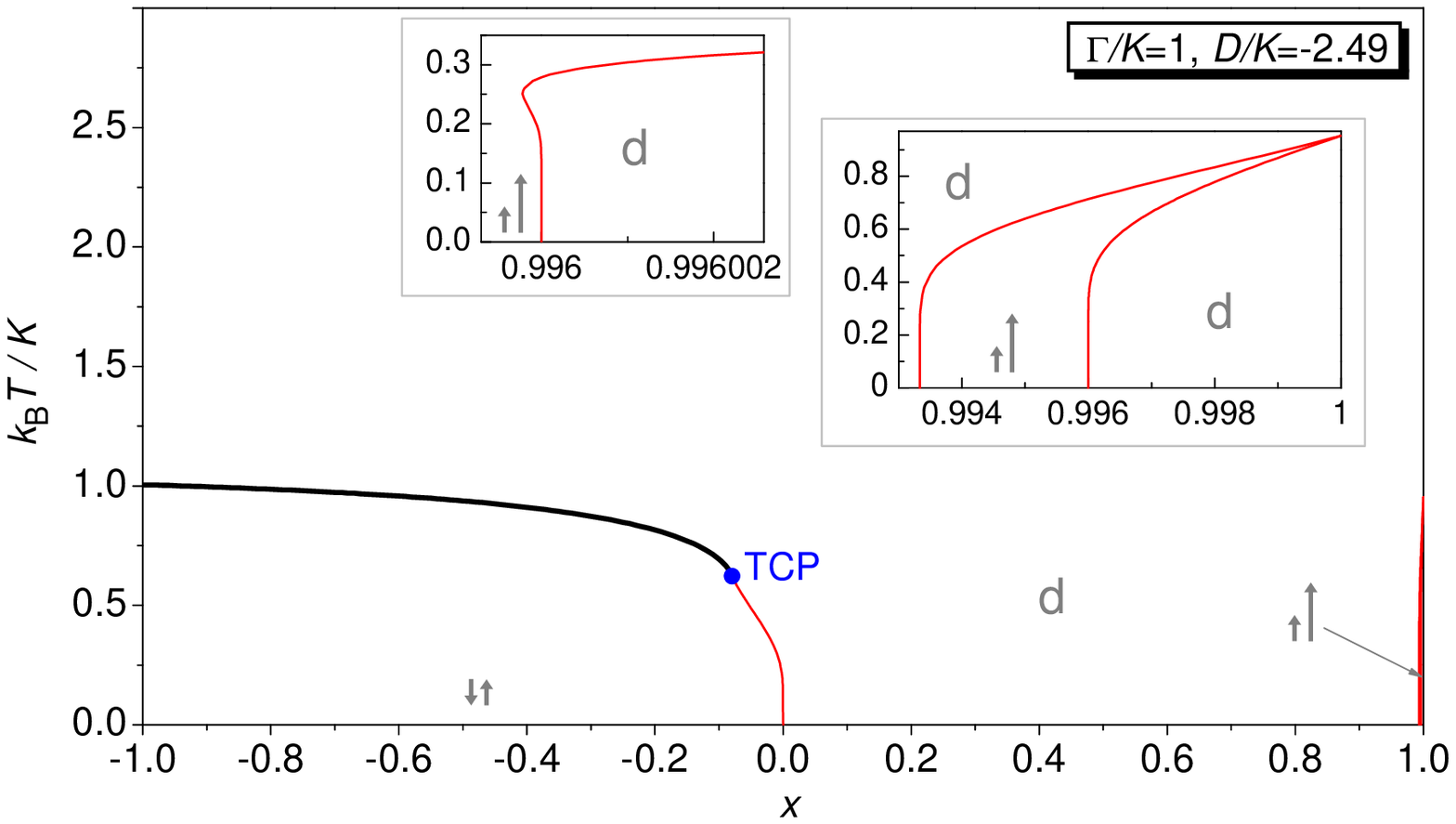}}
\caption{
The same as in Fig. \ref{fig03}, but $D/K=-2.49$ and $\Gamma /K=1$.
} \label{fig12}
\end{figure}

\begin{figure}[h]
\centerline{\includegraphics[width=0.62\textwidth]{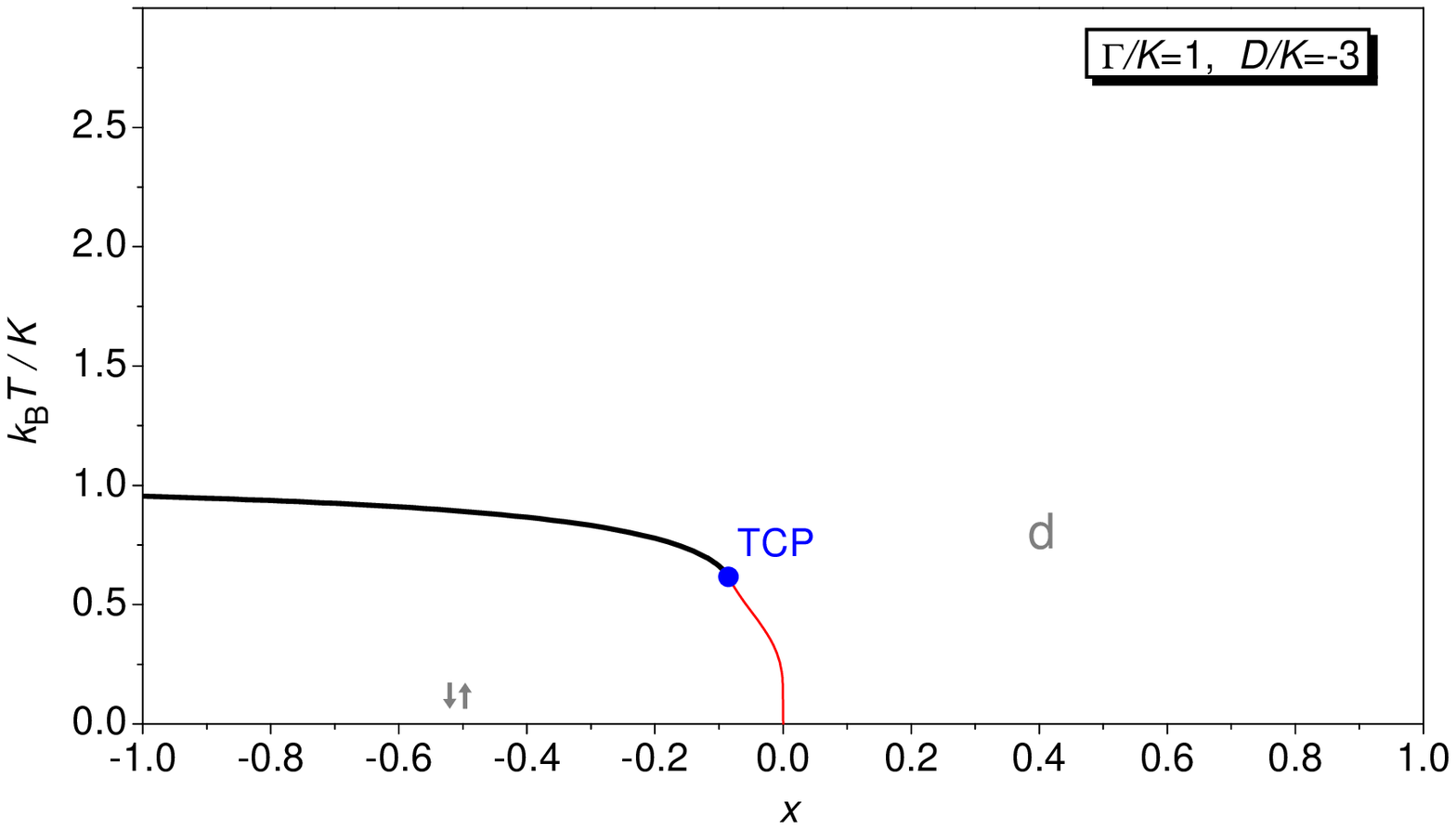}}
\caption{
The same as in Fig. \ref{fig03}, but $D/K=-3$ and $\Gamma /K=1$.
} \label{fig13}
\end{figure}

At first, changes in topologies of $T$ vs $x$ phase diagrams
at variation of the single-ion anisotropy are studied for the
case $\Gamma/K=1$.
Representative $T$ versus $x$ phase diagram cross-sections for different
values of the single-ion anisotropy $D$ of model (1) at $\Gamma/K=1$
are presented in Figs.~\ref{fig01}--\ref{fig13}.

If the value of the single-ion anisotropy is large, the topology of the $(x,
T)$ phase diagrams within MFA is the same as of the diagram given
in Fig. \ref{fig01} (where $D/K=20$). There are the second order and the
first order transition lines separating the disordered phase from
the antiferrimagnetic phase {\bf ai}$_{-3/2,+3/2}$. These lines join
together at the tricritical point (TCP).

In the case $D/K=-1.65$ (see Fig. \ref{fig02})
the phase diagram has topology with a tricritical point, a triple point (TP)
and a critical point (CP) inside the ordered phase. In this case the MFA predicts
two antiferrimagnetic phases {\bf ai}$_{-3/2,+3/2}$, and {\bf ai}$_{-1/2,+3/2}$.
The phase transition {\bf ai}$_{-3/2,+3/2}$~\ra {\bf d} can be of the first or the
second order, but the phase transitions {\bf ai}$_{-3/2,+3/2}$~\ra {\bf ai}$_{-1/2,+3/2}$
and {\bf ai}$_{-1/2,+3/2}$~\ra {\bf d} are of the first order only.

When $D/K=-1.665$ there are three different ordered phases in the system:
two antiferrimagnetic phases {\bf ai}$_{-3/2,+3/2}$, {\bf ai}$_{-1/2,+3/2}$ and
the ferrimagnetic one {\bf i}$_{1/2,+3/2}$.
The phase diagram (see Fig. \ref{fig03})
has the topology with two tricritical points, two triple points, a critical
end point (CEP), a critical point inside the ordered phase, and a re-entrant
region. The sequence of temperature phase transitions
{\bf d}~\ra {\bf i}$_{+1/2,+3/2}$~\ra {\bf d} is possible (at $x \in [0.666112,0.6661157]$).

The phase diagram presented in Fig. \ref{fig04} ($D/K=-1.67$)
has topology with two tricritical points, a critical end point,
a critical point inside the ordered phase, and a double re-entrant region.
At $x \in [0.66799999,0.668]$
the cascade of phase transitions
{\bf i}$_{+1/2,+3/2}$~\ra {\bf d}~\ra {\bf i}$_{+1/2,+3/2}$~\ra {\bf d} is possible.
At $x \in [0.668,0.66809]$ the system undergoes the sequence of temperature
phase transitions {\bf d}~\ra {\bf i}$_{+1/2,+3/2}$~\ra {\bf d}.

In the case $D/K=-1.8$ the $x$ vs temperature phase diagram contains a
tricritical point and two critical points inside the ordered phase
(see Fig. \ref{fig05}).

Topologies of $(x, T)$ phase diagrams at $D/K=-1.879$
and $D/K=-1.89$ shown in Figs. \ref{fig06a} and \ref{fig06}
differ from the described above ones
by the presence of two additional tricritical points. Moreover
at $D/K=-1.89$ the first order phase
transition line has a minimum at $x \approx 0.59$ and the
line of critical points (the second order phase transition line)
has a maximum at $x \approx 0.64$.

At $D/K=-1.95$ the phase diagram
contains three tricritical points (as well as in the case $D/K=-1.89$),
only one critical point inside the ordered phase and a triple point
(instead of the second critical point; see Figs. \ref{fig06} and \ref{fig07}).

In the case $D/K=-1.97$ (see Fig. \ref{fig08})
comparing to the described above cases there is an additional ordered phase:
antiferrimagnetic {\bf ai}$_{-1/2,+1/2}$ phase.
The $(x, T)$ phase diagram at $D/K=-1.97$
has topology with three tricritical points, a triple point and two critical
point inside the ordered phase.

The phase diagram at $D/K=-1.99$ shown in Fig. \ref{fig09}
has topology with three tricritical points, two triple points,
a critical point inside the ordered phase and a re-entrant region.
The cascade of temperature phase transitions
{\bf ai}$_{-1/2,+1/2}$~\ra {\bf ai}$_{-1/2,+3/2}$~\ra {\bf ai}$_{-1/2,+1/2}$
is possible at $x \in [-0.03651,0.0336]$.

In the case $D/K=-1.9999$ (see Fig. \ref{fig10}) the phase diagram
has topology with three tricritical points, three triple points and
two re-entrant regions.
At $x \in [-0.0002115,0.0002015]$ and at $x \in [0.0000213,0.0000415]$
sequences of temperature phase transitions
{\bf ai}$_{-1/2,+1/2}$~\ra {\bf ai}$_{-1/2,+3/2}$~\ra {\bf ai}$_{-1/2,+1/2}$ and
{\bf d}~\ra {\bf ai}$_{-1/2,+1/2}$~\ra {\bf d} are possible, respectively.

At $D/K=-2.1$, $-2.4$ and $-2.49$
there are two different ordered phases in the system:
antiferrimagnetic {\bf ai}$_{-1/2,+1/2}$ and
ferrimagnetic {\bf i}$_{+1/2,+3/2}$.
The $(x,T)$ phase diagrams presented in Figs. \ref{fig11} and \ref{fig11b}
contain three tricritical points while the diagram presented
in Fig. \ref{fig12} contains only one tricritical point.
When $D/K=-2.4$ and $-2.49$  the MFA predicts also the
double re-entrant temperature phase transitions
{\bf i}$_{+1/2,+3/2}$~\ra {\bf d}~\ra {\bf i}$_{+1/2,+3/2}$~\ra {\bf d}
at $x \in [0.95994,0.96]$ and $x \in [0.995999,0.996]$, respectively.
The re-entrant transitions {\bf d}~\ra {\bf i}$_{+1/2,+3/2}$~\ra {\bf d} occur for
$D/K=-2.4$ at $x \in ]0.96,0.96055]$
and for $D/K=-2.49$ at $x \in ]0.996,1[$.

At a small value of the single-ion anisotropy, the topology of the $(x,T)$
phase diagrams within MFA is the same as of the diagram given
in Fig. \ref{fig13} (when $D/K=-3$).
Disordered and antiferrimagnetic {\bf ai}$_{-1/2,+1/2}$ phases
are separated by the lines of phase transition both of the
first and the second orders.

It is obvious, that the described above sequence of different topologies of
($x,T$) phase diagrams under change of the single-ion anisotropy takes place
only at $\Gamma \approx 1$. Other values of the longitudinal field give
essentially different pictures as follows from our investigation of
topologies of ($x,T$) phase diagrams with variation of the longitudinal
field at a fixed value of the single-ion anisotropy.
The most interesting phase diagrams at $D/K=-1.5$ and different
values of $\Gamma$ are presented below. They demonstrate principal changes of
topologies of phase diagrams caused by the field variation at
$D/K \approx -1.5$.

\begin{figure}[h]
\centerline{\includegraphics[width=0.62\textwidth]{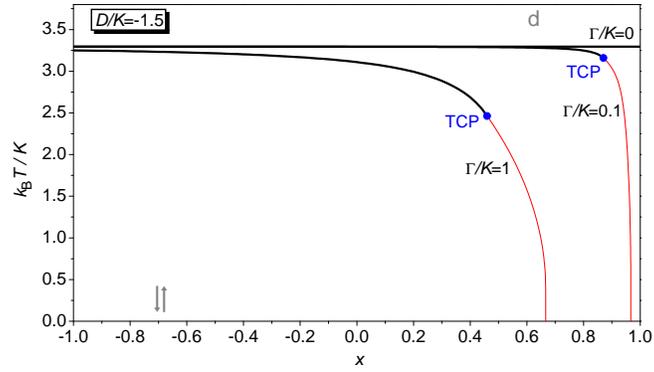}}
\caption{
The same as in Fig. \ref{fig03}, but $D/K=-1.5$ and $\Gamma /K=0.0, \; 0.1, \; 1.0$.
} \label{fig14}
\end{figure}

In the case of the zero longitudinal field,
the MFA results for thermodynamical characteristics
of model (\ref{f1}) do not depend on $x$ \cite{bor1}.
At $k_{\rm B}T/K=3.296$ and any $x \in ]-1,1[$
(see Fig. \ref{fig14}) the system undergoes the second order
temperature phase transition from the antiferromagnetic {\bf af}$_{-3/2,+3/2}$ phase
\cite{Bakchich,bor1} to the paramagnetic phase.

If the value of the longitudinal field is
small, the topology of the $(x,T)$ phase diagram is the
same as of the diagrams given in Fig. \ref{fig14} for $\Gamma/K=0.1$ and $\Gamma/K=1$.
Thick and thin solid lines in these phase diagrams  represent, respectively,
the second order and the first order transition lines from the antiferrimagnetic
{\bf ai}$_{-3/2,+3/2}$ phase to the disordered phase.
It should be noted that the topology of two phase diagrams described above
is the same as the topology of the diagram presented in Fig. \ref{fig01}
in the first part of our study (where the influence of
single-ion anisotropy on $(x,T)$ diagrams is investigated).

\begin{figure}[h]
\centerline{\includegraphics[width=0.62\textwidth]{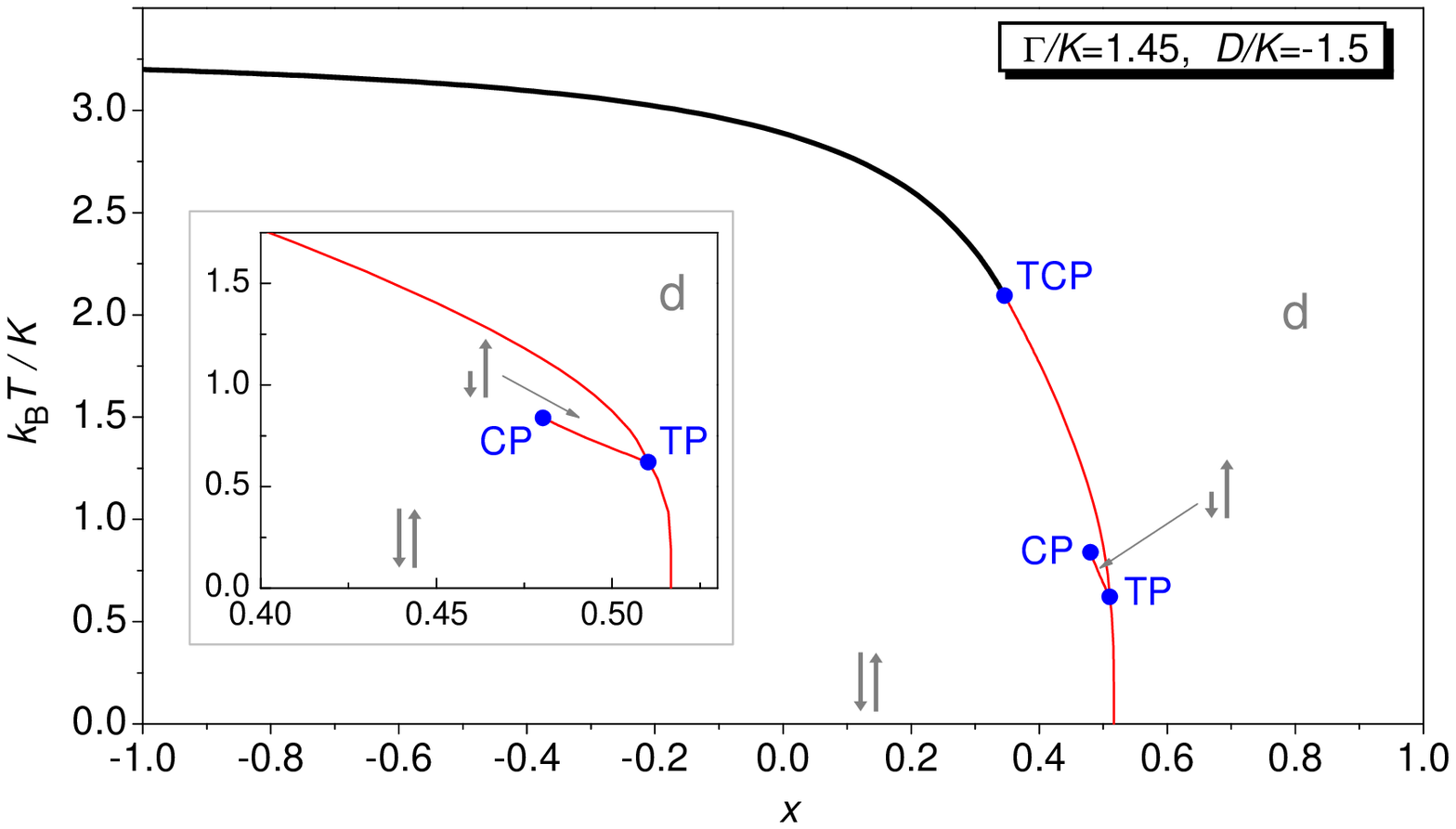}}
\caption{
The same as in Fig. \ref{fig03}, but $D/K=-1.5$ and $\Gamma /K=1.45$.
} \label{fig15}
\end{figure}

In the case $\Gamma/K=1.45$ (see Fig. \ref{fig15})
the phase diagram contains tricritical,  triple and  critical points,
similarly to the case presented in Fig. \ref{fig02}.

\begin{figure}[h]
\centerline{\includegraphics[width=0.62\textwidth]{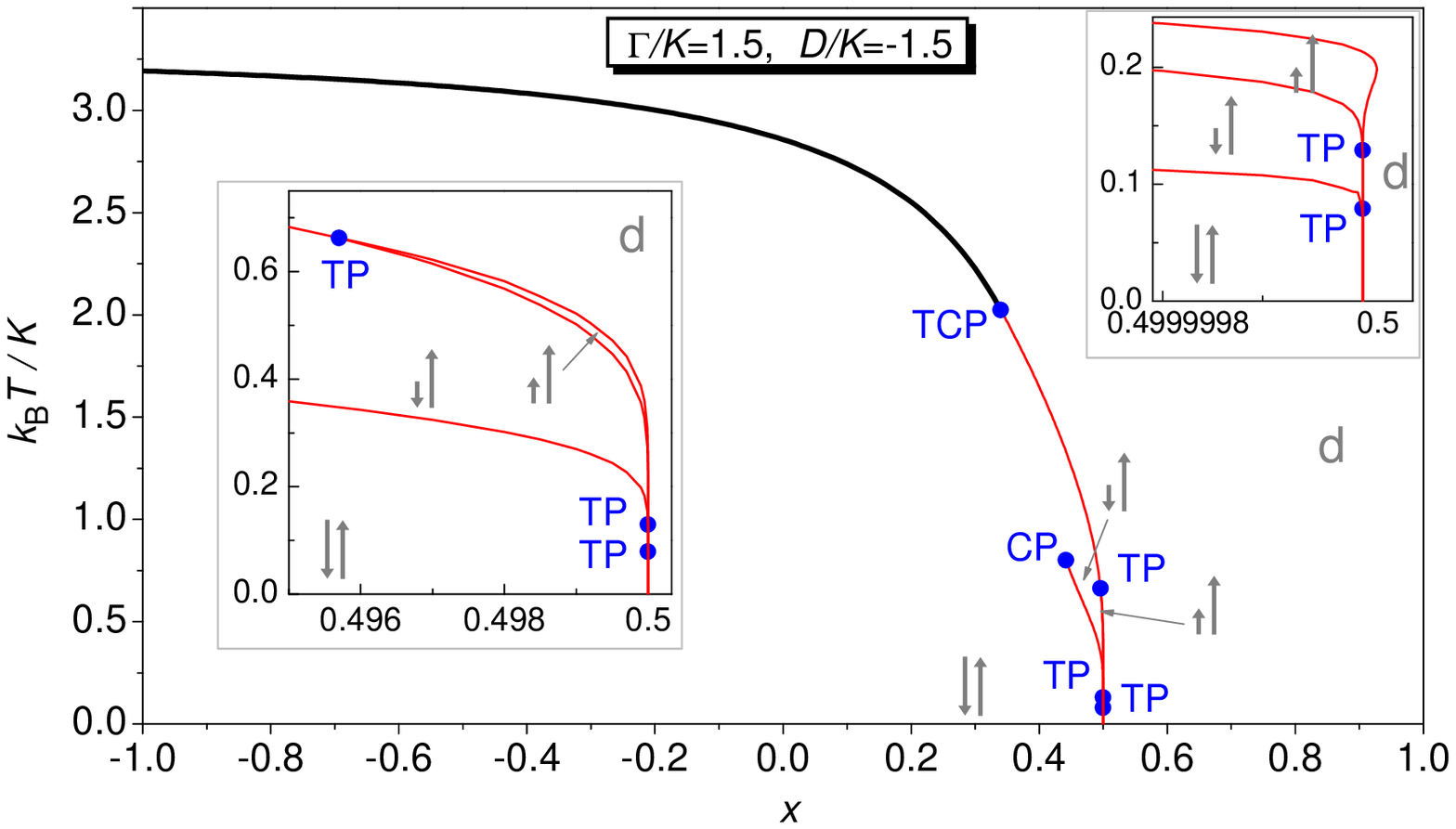}}
\caption{
The same as in Fig. \ref{fig03}, but $D/K=-1.5$ and $\Gamma /K=1.5$.
} \label{fig16}
\end{figure}

The $(x,T)$ phase diagram  at $\Gamma/K=1.5$ (Fig. \ref{fig16}) has
topology with a tricritical point, a critical point inside the ordered phase, three
triple points and a re-entrant region. The cascade of temperature phase transitions
{\bf d}~\ra {\bf i}$_{+1/2,+3/2}$~\ra {\bf d} is possible
(at $x \in [0.5,0.500000014]$).

\begin{figure}[h]
\centerline{\includegraphics[width=0.62\textwidth]{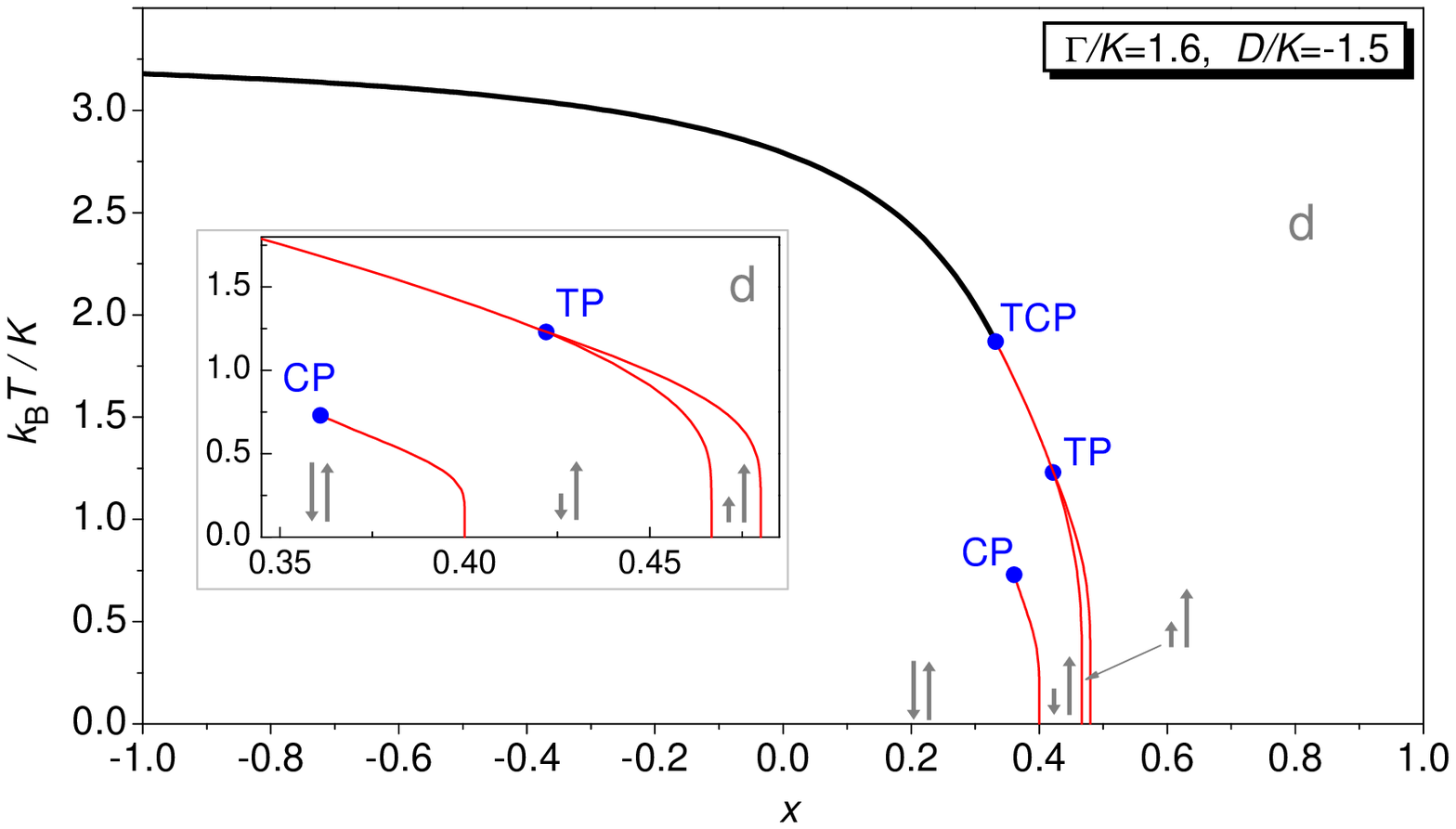}}
\caption{
The same as in Fig. \ref{fig03}, but $D/K=-1.5$ and $\Gamma /K=1.6$.
} \label{fig17}
\end{figure}

At $\Gamma/K=1.6$ (Fig. \ref{fig17}) the phase diagram contains also
a tricritical point, a critical point inside the ordered phase, but only one
triple point in comparison with three TPs in the described above case.

It should be noted that there is a certain similarity between diagrams presented
in Figs. \ref{fig03}, \ref{fig04} and in Figs. \ref{fig16}, \ref{fig17}, respectively.
The diagrams shown in Figs. \ref{fig16} and \ref{fig17} contain
only single tricritical points. Moreover, they contain
triple points instead of critical end points present in
Figs. \ref{fig03} and \ref{fig04}.

\begin{figure}[h]
\centerline{\includegraphics[width=0.62\textwidth]{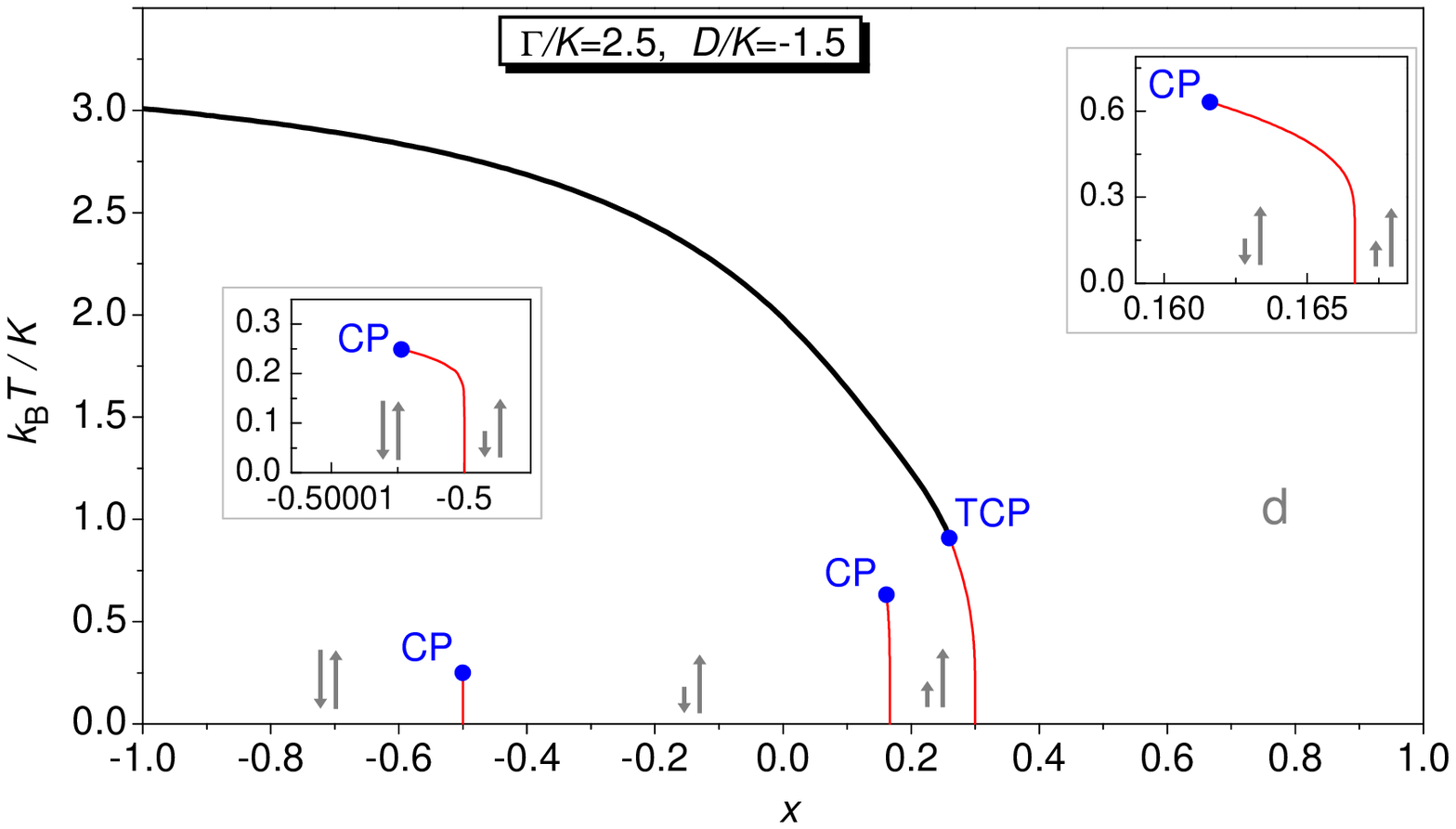}}
\caption{
The same as in Fig. \ref{fig03}, but $D/K=-1.5$ and $\Gamma /K=2.5$.
} \label{fig18}
\end{figure}

In the case $\Gamma/K=2.5$ (see Fig. \ref{fig18}),
similarly to the case presented in Fig. \ref{fig05},
the phase diagram contains two critical points inside the ordered phase
and a single tricritical point.

\begin{figure}[h]
\centerline{\includegraphics[width=0.62\textwidth]{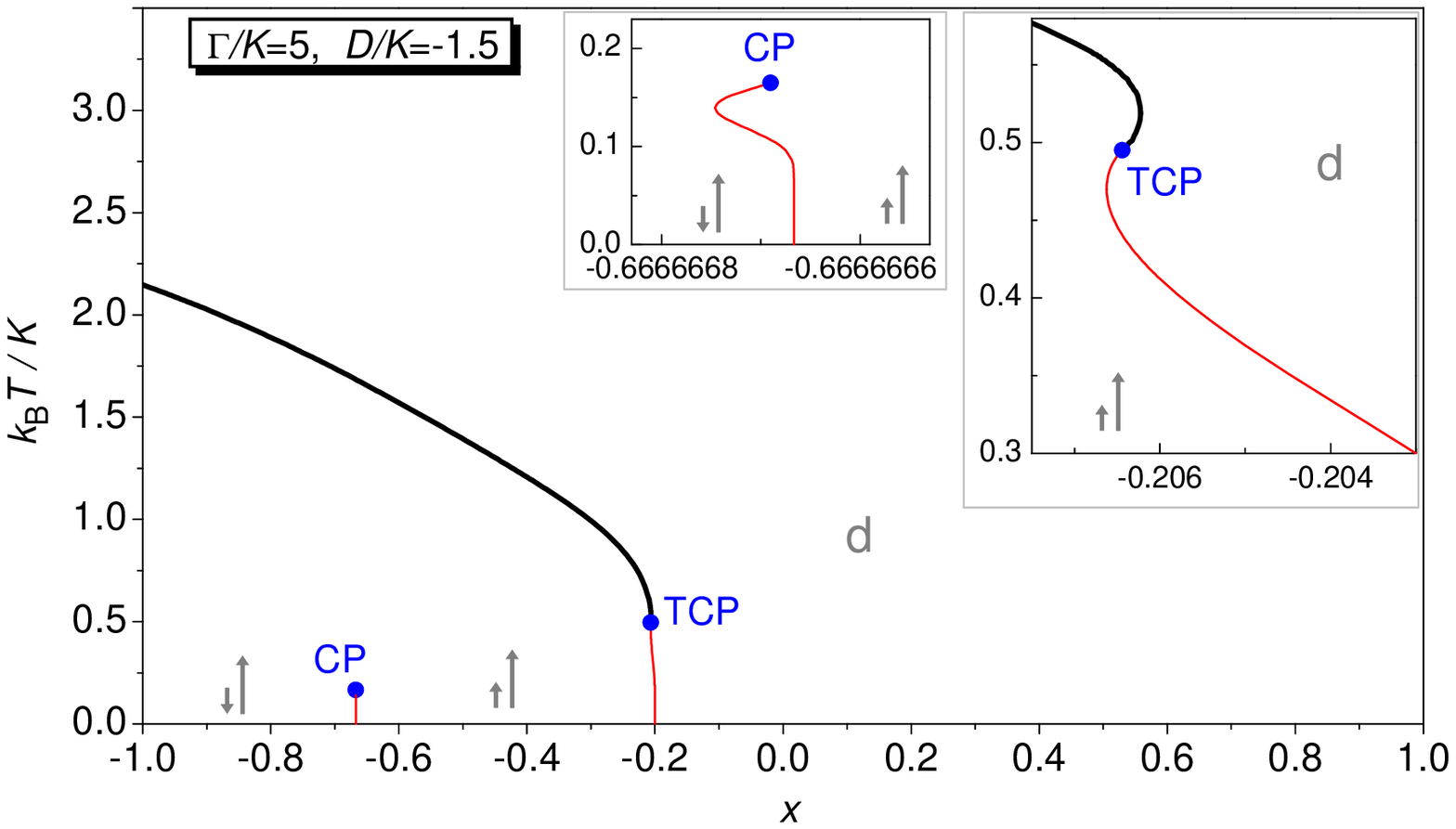}}
\caption{
The same as in Fig. \ref{fig03}, but $D/K=-1.5$ and $\Gamma /K=5$.
} \label{fig19}
\end{figure}

The $T$ vs $x$ phase diagram at $\Gamma/K=5$ (Fig. \ref{fig19})
has topology with a tricritical point, only one critical point,
re-entrant and double re-entrant regions. At $x \in
[-0.66666675,-0.66666669]$ and $x \in [-0.20662,-0.20623]$ the
cascades of temperature phase transitions {\bf ai}$_{-1/2,+3/2}$~\ra
{\bf i}$_{+1/2,+3/2}$~\ra {\bf ai}$_{-1/2,+3/2}$ and {\bf i}$_{+1/2,+3/2}$~\ra {\bf d}~\ra
{\bf i}$_{+1/2,+3/2}$~\ra {\bf d} are possible, respectively. The middle phase
transition in the sequence {\bf i}$_{+1/2,+3/2}$~\ra {\bf d}~\ra
{\bf i}$_{+1/2,+3/2}$~\ra {\bf d}  can be of the first or the second
order.

\begin{figure}[h]
\centerline{\includegraphics[width=0.62\textwidth]{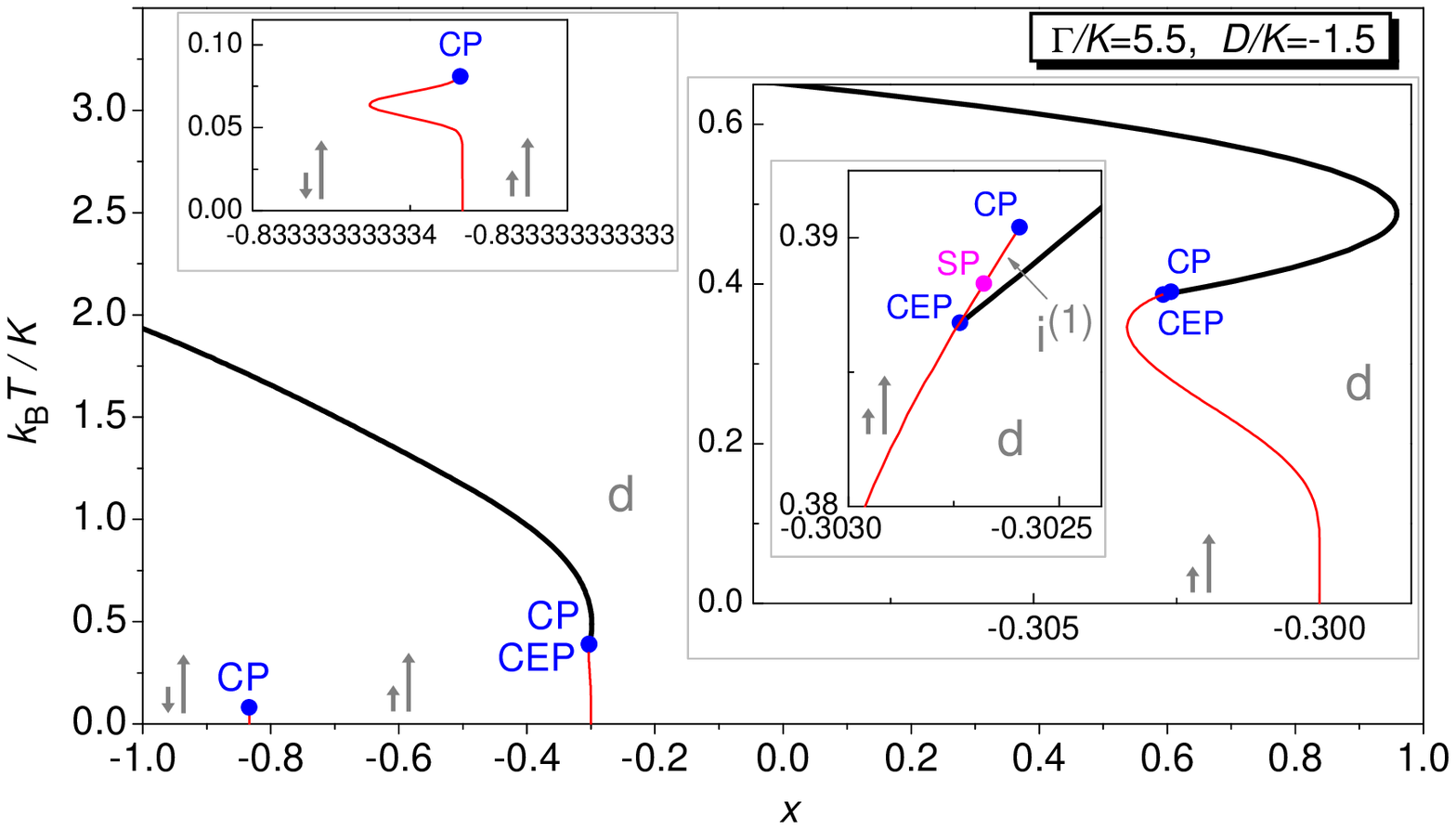}}
\caption{
The $T$ vs $x$ phase diagram at $D/K=-1.5$ and $\Gamma /K=5.5$.
A thick line indicates the
phase transition between ordered and disordered phases of the second order.
Thin lines indicate the first order phase
transitions between ordered and disordered phases
as well as between different ordered phases.
The special points are a critical end point, critical points
and SP.
} \label{fig20}
\end{figure}

The topology of the diagram at $\Gamma/K=5.5$ shown in Fig. \ref{fig20} differs
from the described above one by the fact that the tricritical point
(see Fig. \ref{fig19}) splits into a critical
end point and a critical point inside the ordered phase.
Moreover, there are two additional ferrimagnetic phases
with  sublattice magnetizations which correspond to branches
(solutions of the equation set \qref{f5}) starting from the non-zero
temperature. As mentioned above, such phases are simply designated as
ferrimagnetic ({\bf i}$^{(1)}$ and {\bf i}$^{(2)}$) without indices describing ground
state magnetizations.

Besides the above mentioned special points
(two CP and one CEP) on the phase diagram there is another ``special''
point (SP) at $x_{\rm SP}=-0.302678$, located between
CEP (at $x_{\rm CEP}=-0.302735$) and CP (at $x_{\rm CP}=-0.302594$)
in the curve of the first order phase transition.
The type of the temperature phase transition changes from
{\bf i}$^{(1)}$~\ra {\bf i}$_{+1/2,+3/2}$ to {\bf i}$^{(1)}$~\ra {\bf i}$^{(2)}$ at this SP.
For example, at $x=-0.3027$ the system undergoes the phase
transitions {\bf i}$_{+1/2,+3/2}$~\ra {\bf d}~\ra {\bf i}$^{(1)}$~\ra {\bf i}$_{+1/2,+3/2}$~\ra {\bf d} at
increase of temperature; but at $x=-0.3026$
the system undergoes the
transitions {\bf i}$_{+1/2,+3/2}$~\ra {\bf d}~\ra {\bf i}$^{(1)}$~\ra {\bf i}$^{(2)}$~\ra {\bf d}.

It should be pointed out that the phase diagram in Fig. \ref{fig20} is
complete. The above mentioned SP does not originate any additional curve
of phase transition, what can be proved supplementing the study of temperature
dependencies at fixed $x$ by the investigation of $x$-dependencies of
$m_A$ and $m_B$ at fixed temperatures.
Variation of $x$ in the vicinity of $x_{\rm SP}$  does not indicate any
phase transition at $T=T_{\rm SP}+\varepsilon$ ($\varepsilon~\to 0$).

Such a definition of SP as in Fig. \ref{fig20} is useful for our study of
temperature dependencies of magnetizations and the adopted above classification of
phases.

At $x$ larger then $x_{\rm CP}=-0.302594$ the difference between phases
{\bf i}$^{(1)}$ and {\bf i}$^{(2)}$ disappear as it usually happens above the critical
point. This combined phase is designated as {\bf i}$^{(1|2)}$.
Thus, for example, at $x=-0.301$ the system undergoes the phase
transitions {\bf i}$_{+1/2,+3/2}$~\ra {\bf d}~\ra {\bf i}$^{(1|2)}$~\ra {\bf d} at increase of
temperature.

\begin{figure}[h]
\centerline{\includegraphics[width=0.62\textwidth]{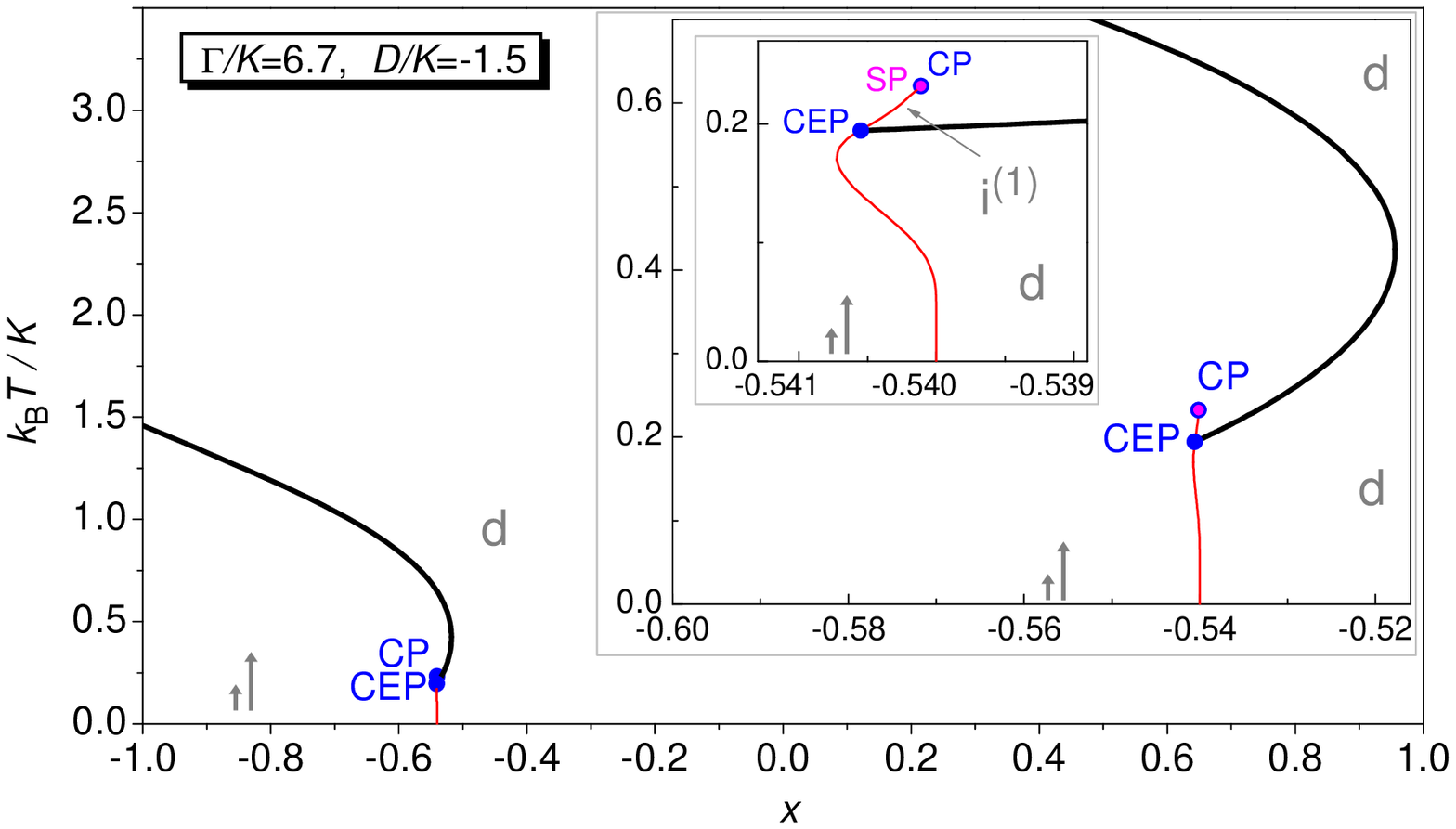}}
\caption{
The same as in Fig. \ref{fig20}, but $D/K=-1.5$ and $\Gamma /K=6.7$.
} \label{fig21}
\end{figure}

\begin{figure}[h]
\centerline{\includegraphics[width=0.62\textwidth]{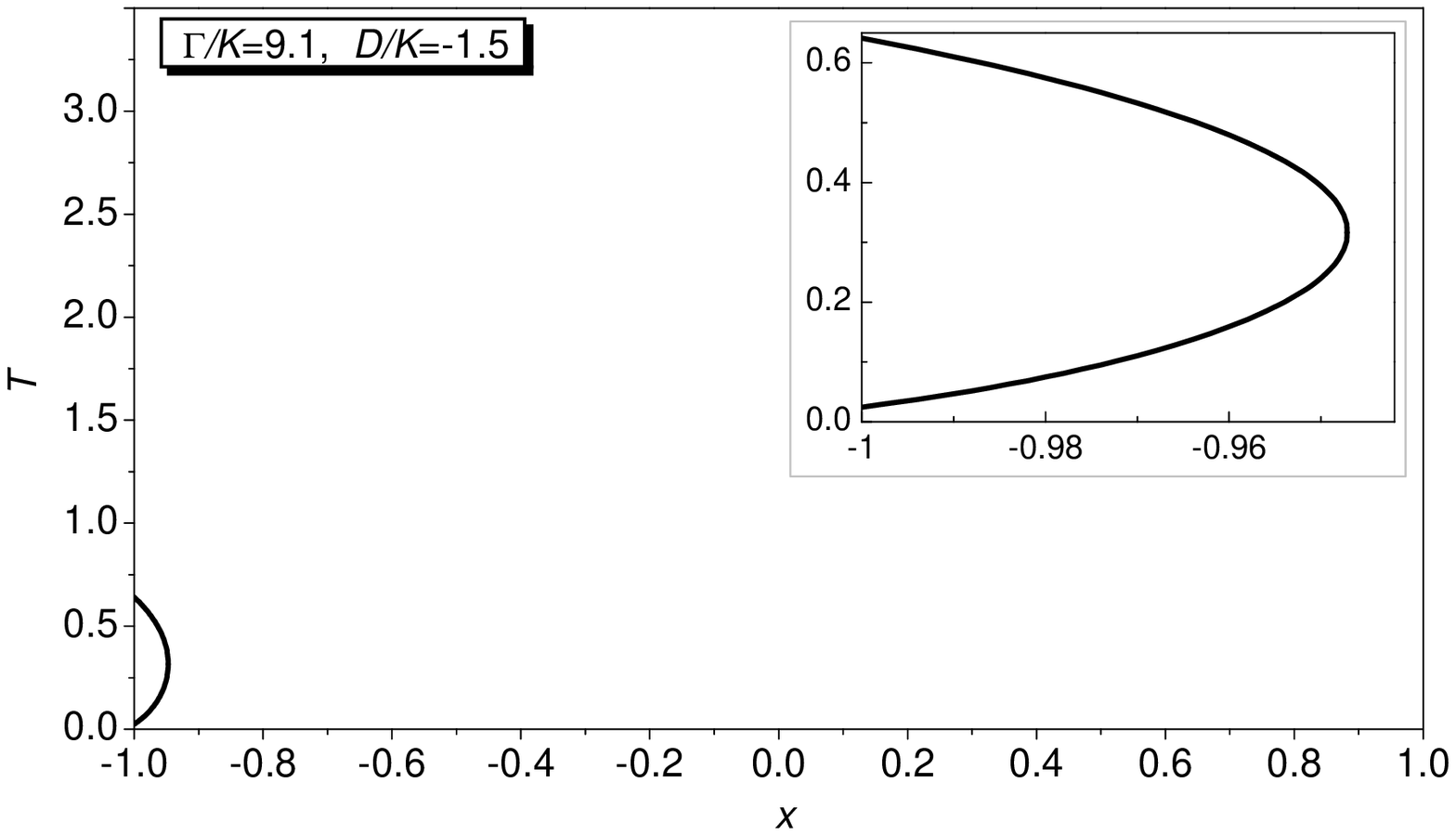}}
\caption{
The same as in Fig. \ref{fig20}, but $D/K=-1.5$ and $\Gamma /K=9.1$.
} \label{fig22}
\end{figure}

Increase of the longitudinal field shifts curves of phase transition in
$(x,T)$ phase diagrams to the left
(see Figs. \ref{fig20} - \ref{fig22}).
For example, the phase {\bf i}$_{-1/2,+3/2}$ is already absent in the phase diagram
at $\Gamma/K=6.7$ (Fig. \ref{fig21}) as compared with the phase diagram at
$\Gamma/K=5.5$ (Fig. \ref{fig20}) and there are only two phases
(disordered and {\bf i}$^{(1|2)}$) in the phase diagram at
$\Gamma/K=9.1$ (Fig. \ref{fig22}).

It should be mentioned, that SP at $x_{\rm SP}=-0.5401103$
almost coincides with the critical point at $x_{\rm CP}=-0.5401101$
in the $(x,T)$ phase diagram at $\Gamma/K=6.7$.
At $x<-0.540725$ (see Fig. \ref{fig21}) only one temperature phase transition
{\bf i}$_{+1/2,+3/2}$~\ra {\bf d} is possible.
At $x \in ]-0.540725, x_{\rm CEP}[$ (where $x_{\rm CEP}=-0.54055$)
the MFA yields the sequence of transitions
{\bf i}$_{+1/2,+3/2}$~\ra {\bf d}~\ra {\bf i}$_{+1/2,+3/2}$~\ra {\bf d}.
At $x \in ]x_{\rm CEP}, x_{\rm SP}[$
the system undergoes the phase
transitions {\bf i}$_{+1/2,+3/2}$~\ra {\bf d}~\ra {\bf i}$^{(1)}$~\ra {\bf i}$_{+1/2,+3/2}$~\ra {\bf d} at
increase of temperature. At $x \in ]x_{\rm SP}, x_{\rm CP}[$
the system undergoes the phase
transitions {\bf i}$_{+1/2,+3/2}$~\ra {\bf d}~\ra {\bf i}$^{(1)}$~\ra {\bf i}$^{(2)}$~\ra {\bf d}.
At $x \in ]x_{\rm CP}, -0.54[$ and at $x \in ]-0.54, -0.51777[$
there are the cascades of temperature phase transitions
{\bf i}$_{+1/2,+3/2}$~\ra {\bf d}~\ra {\bf i}$^{(1|2)}$~\ra {\bf d} and
{\bf d}~\ra {\bf i}$^{(1|2)}$~\ra {\bf d}, respectively.

\section{Conclusions}
The spin-3/2 Blume-Capel model on the rectangular lattice, in which the
short-range
bilinear interactions in perpendicular directions are of opposite signs,
is studied within the mean field approximation. The temperature vs
$x$ phase diagrams ($x$ characterizes coupling between the bilinear
short-range interactions in the perpendicular directions)
at $\Gamma/K=1$ and different values of the single-ion
anisotropy as well as at $D/K=-1.5$ and different values of the longitudinal
magnetic field are obtained.
The phase diagrams presented in this paper illustrate
abundance of the global phase diagram of the considered model.

It has been shown that the usual classification of phases
(like used in works
\cite{Bakchich,Bekhechi,Ekiz2,Keskin1,Keskin2,Baretto,Bakkali,Bakchich2,Xavier})
which distinguishes phases
{\bf ai}$_{-3/2,+3/2}$, {\bf ai}$_{-1/2,+3/2}$, {\bf ai}$_{-1/2,+1/2}$, {\bf i}$_{+1/2,+3/2}$
(where indices correspond to sublattice magnetizations in the ground state)
is mainly suitable for investigation of temperature dependencies of sublattice
magnetizations.
However, a different class of ordered phases is found at the certain values
of model parameters which does not easily fit the mentioned
classification.
These phases exist in the temperature intervals starting from the non-zero
temperatures.

\clearpage

\clearpage

\clearpage

\clearpage

\clearpage

\clearpage


\end{document}